\documentclass[ALICE,manyauthors]{cernphprep}
\usepackage[comma,square,numbers,sort&compress]{natbib}
\usepackage{hyperref}
\usepackage{lineno}
\usepackage{xspace}
\usepackage{xcolor}
\usepackage[T1]{fontenc}
\usepackage{orcidlink} 
\begin{document}
%

\newcommand{\pp}           {pp\xspace}
\newcommand{\ppbar}        {\mbox{$\mathrm {p\overline{p}}$}\xspace}
\newcommand{\XeXe}         {\mbox{Xe--Xe}\xspace}
\newcommand{\PbPb}         {\mbox{Pb--Pb}\xspace}
\newcommand{\pA}           {\mbox{pA}\xspace}
\newcommand{\pPb}          {\mbox{p--Pb}\xspace}
\newcommand{\AuAu}         {\mbox{Au--Au}\xspace}
\newcommand{\dAu}          {\mbox{d--Au}\xspace}

\newcommand{\s}            {\ensuremath{\sqrt{s}}\xspace}
\newcommand{\snn}          {\ensuremath{\sqrt{s_{\mathrm{NN}}}}\xspace}
\newcommand{\pt}           {\ensuremath{p_{\rm T}}\xspace}
\newcommand{\pT}           {\ensuremath{p_{\rm T}}\xspace}
\newcommand{\meanpt}       {$\langle p_{\mathrm{T}}\rangle$\xspace}
\newcommand{\ycms}         {\ensuremath{y_{\rm CMS}}\xspace}
\newcommand{\ylab}         {\ensuremath{y_{\rm lab}}\xspace}
\newcommand{\etarange}[1]  {\mbox{$\left | \eta \right |~<~#1$}}
\newcommand{\yrange}[1]    {\mbox{$\left | y \right |~<~#1$}}
\newcommand{\dndy}         {\ensuremath{\mathrm{d}N_\mathrm{ch}/\mathrm{d}y}\xspace}
\newcommand{\dndeta}       {\ensuremath{\mathrm{d}N_\mathrm{ch}/\mathrm{d}\eta}\xspace}
\newcommand{\avdndeta}     {\ensuremath{\langle\dndeta\rangle}\xspace}
\newcommand{\dNdy}         {\ensuremath{\mathrm{d}N_\mathrm{ch}/\mathrm{d}y}\xspace}
\newcommand{\Npart}        {\ensuremath{N_\mathrm{part}}\xspace}
\newcommand{\Ncoll}        {\ensuremath{N_\mathrm{coll}}\xspace}
\newcommand{\dEdx}         {\ensuremath{\textrm{d}E/\textrm{d}x}\xspace}
\newcommand{\RpPb}         {\ensuremath{R_{\rm pPb}}\xspace}

\newcommand{\nineH}        {$\sqrt{s}~=~0.9$~Te\kern-.1emV\xspace}
\newcommand{\seven}        {$\sqrt{s}~=~7$~Te\kern-.1emV\xspace}
\newcommand{\twoH}         {$\sqrt{s}~=~0.2$~Te\kern-.1emV\xspace}
\newcommand{\twosevensix}  {$\sqrt{s}~=~2.76$~Te\kern-.1emV\xspace}
\newcommand{\five}         {$\sqrt{s}~=~5.02$~Te\kern-.1emV\xspace}
\newcommand{\twosevensixnn}{$\sqrt{s_{\mathrm{NN}}}~=~2.76$~Te\kern-.1emV\xspace}
\newcommand{\fivenn}       {$\sqrt{s_{\mathrm{NN}}}~=~5.02$~Te\kern-.1emV\xspace}
\newcommand{\LT}           {L{\'e}vy-Tsallis\xspace}
\newcommand{\GeVc}         {Ge\kern-.1emV/$c$\xspace}
\newcommand{\MeVc}         {Me\kern-.1emV/$c$\xspace}
\newcommand{\TeV}          {Te\kern-.1emV\xspace}
\newcommand{\GeV}          {Ge\kern-.1emV\xspace}
\newcommand{\MeV}          {Me\kern-.1emV\xspace}
\newcommand{\GeVmass}      {Ge\kern-.2emV/$c^2$\xspace}
\newcommand{\MeVmass}      {Me\kern-.2emV/$c^2$\xspace}
\newcommand{\lumi}         {\ensuremath{\mathcal{L}}\xspace}

\newcommand{\ITS}          {\rm{ITS}\xspace}
\newcommand{\TOF}          {\rm{TOF}\xspace}
\newcommand{\ZDC}          {\rm{ZDC}\xspace}
\newcommand{\ZDCs}         {\rm{ZDCs}\xspace}
\newcommand{\ZNA}          {\rm{ZNA}\xspace}
\newcommand{\ZNC}          {\rm{ZNC}\xspace}
\newcommand{\SPD}          {\rm{SPD}\xspace}
\newcommand{\SDD}          {\rm{SDD}\xspace}
\newcommand{\SSD}          {\rm{SSD}\xspace}
\newcommand{\TPC}          {\rm{TPC}\xspace}
\newcommand{\TRD}          {\rm{TRD}\xspace}
\newcommand{\VZERO}        {\rm{V0}\xspace}
\newcommand{\VZEROA}       {\rm{V0A}\xspace}
\newcommand{\VZEROC}       {\rm{V0C}\xspace}
\newcommand{\Vdecay} 	   {\ensuremath{V^{0}}\xspace}

\newcommand{\ee}           {\ensuremath{e^{+}e^{-}}} 
\newcommand{\pip}          {\ensuremath{\pi^{+}}\xspace}
\newcommand{\pim}          {\ensuremath{\pi^{-}}\xspace}
\newcommand{\kap}          {\ensuremath{\rm{K}^{+}}\xspace}
\newcommand{\kam}          {\ensuremath{\rm{K}^{-}}\xspace}
\newcommand{\pbar}         {\ensuremath{\rm\overline{p}}\xspace}
\newcommand{\kzero}        {\ensuremath{{\rm K}^{0}_{\rm{S}}}\xspace}
\newcommand{\lmb}          {\ensuremath{\Lambda}\xspace}
\newcommand{\almb}         {\ensuremath{\overline{\Lambda}}\xspace}
\newcommand{\Om}           {\ensuremath{\Omega^-}\xspace}
\newcommand{\Mo}           {\ensuremath{\overline{\Omega}^+}\xspace}
\newcommand{\X}            {\ensuremath{\Xi^-}\xspace}
\newcommand{\Ix}           {\ensuremath{\overline{\Xi}^+}\xspace}
\newcommand{\Xis}          {\ensuremath{\Xi^{\pm}}\xspace}
\newcommand{\Oms}          {\ensuremath{\Omega^{\pm}}\xspace}
\newcommand{\degree}       {\ensuremath{^{\rm o}}\xspace}

\graphicspath{{FIGURES/}}
\begin{titlepage}
\PHyear{2022}       
\PHnumber{230}      
\PHdate{27 October}  

\title{Two-particle transverse momentum correlations\\in pp and p--Pb collisions at energies available at the CERN Large Hadron Collider}
\ShortTitle{$G_{2}$ in small systems}   

\Collaboration{ALICE Collaboration\thanks{See Appendix~\ref{app:collab} for the list of collaboration members}}
\ShortAuthor{ALICE Collaboration} 

\begin{abstract}
Two-particle transverse momentum differential correlators, recently measured in Pb--Pb collisions at energies available at the CERN Large Hadron Collider (LHC),  provide   an additional tool  to gain insights into particle production mechanisms and infer transport properties, such as the ratio of shear viscosity to entropy density, of the medium created in Pb--Pb collisions. The longitudinal long-range correlations and the large azimuthal anisotropy measured at low transverse momenta in small collision systems, namely pp and p--Pb, at LHC energies resemble manifestations of collective behaviour. This suggests that locally equilibrated matter may be produced in these small collision systems, similar to what is observed in Pb--Pb collisions. In this work, the same two-particle transverse momentum differential correlators are exploited  in pp and p--Pb collisions at $\sqrt{s} = 7\;\text{TeV}$ and $\sqrt{s_{\rm NN}} = 5.02\;\text{TeV}$, respectively, to seek evidence for viscous effects. Specifically, the strength and shape of the correlators are studied as a function of the produced particle multiplicity to identify evidence for longitudinal broadening that might reveal the presence of viscous effects in these smaller systems. The measured correlators and their evolution from pp and p--Pb to Pb--Pb collisions are additionally compared to predictions from Monte Carlo event generators, and the potential presence of viscous effects is discussed.

\end{abstract}
\end{titlepage}

\setcounter{page}{2} 


\section{Introduction} 

Studies  at  the  Large Hadron Collider (LHC)
and the Relativistic Heavy-Ion Collider (RHIC) have shown that quark--gluon plasma  (QGP) matter is produced in relativistic collisions of large nuclei~\cite{Adams:2005dq,Adcox:2004mh,Arsene20051,Back:2004je,ALICE:2010suc,ALICE:2013osk,ALICE:2015mjv,ALICE:2016flj,CMS:2017uuv,CMS:2018zza,ALICE:2019hno}, and considerable efforts have been undertaken to measure some of the key properties of this phase of matter. One such property, the shear viscosity per unit entropy density, $\eta/s$, has received much attention from both the theoretical and experimental communities~\cite{Song:2010mg,Qiu:2011hf,Song:2012tv,Gale:2013da,Heinz:2013th,Bernhard:2019bmu}. 
Measurements of anisotropic flow coefficients as well as symmetric cumulants, which are correlations between flow coefficients of different orders, in particular, have been successfully exploited to determine the extent to which the QGP is an almost perfect fluid and many advances have been accomplished in improving estimates of the QGP $\eta/s$ based on such  measurements~\cite{ALICE:2021klf,ALICE:2016kpq,Niemi:2015qia}. 

A new line of investigation was recently undertaken to extract values of the QGP $\eta/s$. This approach  is based on measurements of the longitudinal broadening of a specific type of transverse momentum differential two-particle correlation function known as $G_2$ in the recent literature~\cite{Gavin:2006xd,Sharma:2008qr,Gavin:2016hmv}.  While the technique is relatively new and still needs  to be fully vetted by detailed (3+1)-dimensional hydrodynamical calculations, a recent measurement of the longitudinal broadening of $G_2$ in central Pb--Pb collisions, by the ALICE Collaboration~\cite{ALICE:2019smr}, is found to yield   an $\eta/s$  range compatible with estimations based on anisotropic flow~\cite{Gonzalez:2020bqm}. This agreement suggests that the new approach has merits and potential in furthering the understanding of the properties of QGP matter produced in collisions of large nuclei. 

In proton--proton (pp) and proton--lead (p--Pb) collisions, femtoscopy radii, related to the estimated size of the system, and average transverse momentum, $\langle p_{\rm T} \rangle$, increase with the multiplicity of produced particles~\cite{Abelev:2013bla,ALICE:2014xrc,ALICE:2015hav,CMS:2017mdg,Heinz:2019dbd}, which implies that the system lives longer as the multiplicity increases, i.e. as the system size increases. In turn, this means that radial flow would have more time to develop. In these terms, for a given system size, viscous effects, if present, will have a certain time to manifest themselves by transferring momentum between neighbouring fluid cells. This transfer of momentum will make the correlation function for that system size to acquire, at the end, a certain longitudinal width which reflects the reach of the viscous effects. 
With larger multiplicity, the larger the system size is and the larger the time the system lives, the reach of the viscous effects will also be larger, which causes the longitudinal width of the correlation function to enlarge. 
Overall, the correlation function broadens longitudinally with the system size, the system lifetime, and the system multiplicity.
In contrast, as more radial flow builds up with increasing multiplicity, system size, and lifetime, a narrower width is expected in the correlator azimuthal dimension. 
The fact that viscous effects are, in principle, independent of the charge and that radial flow has a strong charge dependent component~\cite{ALICE:2015nuz,ALICE:2018jco} 
implies that the charge independent (CI) and charge dependent (CD) correlators play complementary roles in the evaluation of the interplay between these effects.

It is thus natural to consider whether the above technique could also be exploited in the study of small systems, such as proton--proton (pp) and proton--lead (p--Pb) collisions. Measurements of anisotropic flow coefficients and multi-particle cumulants indicate that strong collective behaviour exists in high-multiplicity pp and p--Pb collisions~\cite{CMS:2010ifv,CMS:2015fgy,ATLAS:2015hzw}. Several calculations based on hydrodynamics models~\cite{Werner:2013tya,Weller:2017tsr} suggest that the observed coefficients can in fact be interpreted  as evidence of collective flow in  high multiplicity pp and p--Pb collisions. An important  question  is whether the apparent collectivity arises  from the production of a medium, albeit much smaller than that produced in Pb--Pb collisions, or is due to other types of correlation sources. 

In this context, it is of interest to examine whether measurements of $G_2$ in small collision systems can be exploited to identify the existence of a droplet sufficiently large and long lived such that viscous effects can yield a longitudinal  broadening of  the correlator. 
A particularly appealing aspect of the $G_2$ correlator is that it can be studied for charge dependent and charge independent pairs of particles. 
The charge independent $G_2$ correlator, hereafter denoted $G_2^{\rm CI}$, is by construction sensitive to momentum current correlations. It is thus expected to exhibit a progressive broadening from small to large multiplicity collisions if these involve a long lived QGP matter undergoing both longitudinal and transverse expansion. In this expansion, shear viscous effects can transform stochastic radial currents and produce a longitudinal broadening of the correlator. 
However, the shape and evolution of $G_2^{\rm CI}$ may also be impacted by the presence of hadronic decays and radial flow and  by minijet correlations resulting from  parton shower evolution, string-breaking, and hadronisation effects. 
Measurements of the  charge dependent $G_2$ correlator, hereafter noted $G_2^{\rm CD}$, can be exploited to specifically study these contributions. 
Indeed,   $G_2^{\rm CD}$ and other two-particle differential correlation functions are found to be particularly sensitive to the presence of hadronic resonance decays and radial flow, but somewhat less sensitive to shear viscous effects~\cite{Pruneau:2007ua,Gavin:2008ta}. They can thus, in principle, be used to assess these effects and ``calibrate them'' out of measurements of $G_2^{\rm CI}$.  

Prior measurements of two-particle azimuthal correlations in the p--Pb system~\cite{ALICE:2012eyl} compared the strength of azimuthal modulations in collisions producing the largest and  lowest charged particle multiplicity. They found evidence of sizable flow-like azimuthal correlation structures in high-multiplicity p--Pb collisions but did not study their  pseudorapidity  dependence in detail.
It is thus the primary goal of this work to extend those correlation studies and measure the evolution of  $G_2^{\rm CD}$ and   $G_2^{\rm CI}$ as a function of the produced charged particle multiplicity in both pp and p--Pb collisions. Then, this will allow seeking evidence for  longitudinal broadening of $G_2^{\rm CI}$ signaling viscous effects that should happen if relatively long lived QGP matter is produced in these collisions.  

This work is organised as follows. Section~\ref{sec:anametho} defines the two-particle correlator $G_2$ and presents the measurement  methodology, while Sec.~\ref{sec:expsetup} describes the experimental details  and  corrections applied to the data. Section~\ref{sec:uncertainties} presents the techniques  used to determine statistical and systematic  uncertainties on the measured correlation function amplitudes  and their characteristics, reported in Sec.~\ref{sec:results}. The method used to characterise the shape of the correlation functions and its evolution with multiplicity is presented in Sec.~\ref{sec:shapeextraction} . Measurements of the evolution of the longitudinal and azimuthal widths of the correlators are compared to model calculations  in Sec.~\ref{sec:models}. A discussion of the results and  models is presented in Sec.~\ref{sec:discussion} followed by a summary of the conclusions of this work in Sec.~\ref{sec:conclusions}.

\section{Analysis methodology}
\label{sec:anametho}

The $G_2$ correlator is designed to be proportional to the magnitude of momentum currents, the transferring of momentum fluctuations, and their correlations, from which viscous effects can be inferred~\cite{Gavin:2006xd,Sharma:2008qr}.
It is defined as
\begin{equation}
G_2\left( \eta_1,\varphi_1,\eta_2,\varphi_2 \right) =  
  \frac{1}{\langle p_{\rm T,1} \rangle \langle p_{\rm T,2} \rangle} \left[ 
  \frac{\int_{\Omega} p_{\rm T, 1}p_{\rm T, 2}\, \rho_2(\vec p_1,\vec p_2) 
      \, {\rm d}\,p_{\rm T, 1}{\rm d}\,p_{\rm T, 2}}
    {\int_{\Omega} \rho_1(\vec p_1)\, {\rm d}\,p_{\rm T, 1}
      \; \int_{\Omega} \rho_1(\vec p_2)\, {\rm d}\,p_{\rm T, 2}} 
    - \langle p_{\rm T,1}\rangle  \langle p_{\rm T,2}\rangle 
    \right]\text{,}
  \label{eq:normG2}
\end{equation}
where  $\rho_{1}(\vec{p}_{\rm i})$ and $\rho_{2}(\vec{p}_{1},\vec{p}_{2})$ represent single-particle   and pair densities  computed as 
\begin{align}
\rho_{1}(\vec{p}_{\rm i})&= \frac{{\rm d}^{3}N}{{\rm d}p_{{\rm T},{\rm i}}\,{\rm d}\eta_{\rm i}\,{\rm d}\varphi_{\rm i}}, \\ 
\rho_{2}(\vec{p}_{1},\vec{p}_{2}) &= \frac{{\rm d}^{6}N}{{\rm d}p_{{\rm T},1}\,{\rm d}\eta_{1}\,{\rm d}\varphi_{1}\,{\rm d}p_{{\rm T},2}\,{\rm d}\eta_{2}\,{\rm d}\varphi_{2}},
\end{align}
 with  particle three-momenta $\vec p_{\rm i}=(\eta_{\rm i},\varphi_{\rm i},p_{\rm T,i})$  and components $\eta_{\rm i},\varphi_{\rm i},p_{\rm T,i}$ corresponding to the   pseudorapidity,  azimuthal angle, and  transverse momentum
 of particles ${\rm i}=1,2$, composing pairs. Transverse momentum averages $\langle p_{{\rm T},{\rm i}} \rangle$  are calculated according to
\begin{equation}
\langle p_{{\rm T},{\rm i}} \rangle(\eta_{\rm i},\varphi_{\rm i}) =
\frac{
\int_{\Omega} \rho_{1}(\eta_{\rm i},\varphi_{\rm i},p_{\rm T,i}) p_{\rm T,i} {\rm d}p_{\rm T,i}
}
{
\int_{\Omega} \rho_{1}(\eta_{\rm i},\varphi_{\rm i},p_{\rm T,i}) {\rm d}p_{\rm T,i}
}.
\end{equation}
Integrals are computed in  the measurement acceptance $\Omega$.
Measurements of $G_2(\eta_1,\varphi_1,\eta_2,\varphi_2)$ are averaged across the longitudinal and azimuthal acceptances in which the measurement is performed to obtain $G_2(\Delta \eta, \Delta \varphi)$, where $\Delta \eta = \eta_1-\eta_2$ and $\Delta \varphi = \varphi_1-\varphi_2$, with a procedure similar to that used for the two-particle number correlator $R_2$ and the two-particle transverse momentum correlator $P_2$~\cite{ALICE:2018jco}, as well as for  measurements of $G_2$ in Pb--Pb collisions~\cite{ALICE:2019smr}.

In order to account  for distinct efficiency losses associated with positively $(+)$ and negatively $(-)$ charged particles, $G_2$ correlators are first measured for pairs of $(++)$, $(--)$, $(-+)$, and $(+-)$
charged hadrons. These measurements are combined to yield  like-sign (LS) and unlike-sign (US) pairs correlators  $G_2^{\rm LS}= \frac{1}{2}(G_2^{++}+G_2^{--})$ and $G_2^{\rm US}= \frac{1}{2}(G_2^{+-}+G_2^{-+})$. In turn, these are further combined  to obtain the charge dependent and the charge independent correlators  defined as  $G_2^{\rm{CD}} = \frac{1}{2} \left(G_2^{\rm{US}}- G_2^{\rm{LS}}\right) $ and $G_2^{\rm{CI}} = \frac{1}{2} \left(G_2^{\rm{US}} + G_2^{\rm{LS}}\right) $, respectively~\cite{ALICE:2018jco}. 

The $G_2^{\rm{CD}}$ and $G_2^{\rm{CI}}$ correlators are measured in pp and p--Pb collisions using  event classes based on the average charged particle multiplicity detected at forward pseudorapidities. 
The multiplicity evolution of the shape and strength of these correlators is then extracted and analysed as described in Sec.~\ref{sec:shapeextraction}.

\section{Datasets and experimental setup}
\label{sec:expsetup}

The results presented in this article are based on $6.4 \times 10^7$ selected minimum bias (MB) pp collisions at centre-of-mass energy $\sqrt{s}=7\;\text{TeV}$ and $5.4 \times 10^7$ selected MB p--Pb collisions at centre-of-mass energy per nucleon--nucleon collision $\sqrt{s_{\rm NN}}=5.02\;\text{TeV}$ collected during the 2010 and 2013 LHC runs, respectively, with the ALICE detector. 
Detailed descriptions of the ALICE subsystems and their respective performance are given in Refs.~\cite{ALICE:2008ngc,Abelev:2014ffa}. 

The MB trigger was configured to provide a high efficiency for hadronic events. It required coincident signals in the V0A and V0C  scintillator arrays~\cite{ALICE:2013axi}, covering the pseudorapidity ranges $2.8 < \eta < 5.1$ and $-3.7 <\eta < -1.7$, respectively. 
Calibrated SPD and V0 signal amplitudes were used to estimate the charged particle multiplicity production in these pseudorapidity ranges. The resulting multiplicity distribution was used to establish nine multiplicity classes corresponding to 0--5\% (highest multiplicity), 5--10\%, 10--20\%, 20--30\%, 30--40\%, 40--50\%, 50--60\%, 60--70\%, and  70--80\% (lowest multiplicity) of the inelastic cross section. 
The correlators $G_2^{\rm CI}$ and $G_2^{\rm CD}$  are extracted 
independently in each of these multiplicity classes and the  evolution of their widths is reported as a function of the average number of charged particles $[ N_{\rm ch}]$ measured in the fiducial acceptance of the measurement.

The collision vertex position of each event, called primary vertex (PV), is determined from the charged particle tracks reconstructed in the Inner Tracking System (ITS) and the Time Projection Chamber (TPC). Only events with a reconstructed PV position within 7 cm from the nominal interaction point along the beam direction were included in the analysis. 
Background events from beam interactions with residual gas in the beam pipe are removed using the timing information in the V0. Pileup events having multiple interaction vertices are discarded based on information from the Silicon Pixel Detector (SPD) constituting the  two inner layers of the ITS. 
Extra activity in slow response detectors (e.g., TPC) relative to that in fast detectors (e.g., V0A and V0C scintillators) resulting largely from   out of bunch pileup events, is additionally used to discard these events. 

Charged particle tracks are reconstructed using the ITS and TPC detectors and required to have transverse momenta and pseudorapidities within the ranges $0.2\le \pt \le 2.0$ \GeVc and  $|\eta|<0.8$, respectively. 
Good track quality is assured by retaining only tracks with more than 70 reconstructed TPC space points, out of a maximum of 159,  for the  analysis. 
A criterion on the maximum distance of closest approach (DCA) to the reconstructed PV of less than $2\;\text{cm}$ in the longitudinal dimension and a $p_{\rm T}$-dependent maximum DCA in the transverse direction, ranging from $0.20\;\text{cm}$ at $p_{\rm T} = 0.2\;\text{GeV}/c$ down to $0.036\;\text{cm}$ at $p_{\rm T} = 2\;\text{GeV}/c$ for pp collisions and from $0.22\;\text{cm}$ at $p_{\rm T} = 0.2\;\text{GeV}/c$ down to $0.031\;\text{cm}$ at $p_{\rm T} = 2\;\text{GeV}/c$ for p--Pb collisions, is applied to minimise contamination by secondary tracks. Moreover,
electrons (positrons), which originate mainly from photon conversions into $\rm e^{+}e^{-}$ pairs, are suppressed by removing tracks with a specific energy loss in the TPC gas, ${\rm d} E/{\rm d} x$,   within three standard deviations, $3\sigma_{{\rm d} E/{\rm d} x}$, of the expected value for electrons and more than  $5\sigma_{{\rm d} E/{\rm d} x}$ away from the $\pi$ and K expectation values.

Corrections for single track losses due to detector non-uniformity are based on a weighting technique~\cite{Ravan:2013lwa}. 
Weights are  calculated separately for positive and negative tracks as a function of $\eta$, $\varphi$, and $\pt$ and averaged across the measured ranges of multiplicity and primary vertex position. 
Weights are used to flatten the track yield in both pseudorapidity and in azimuth for the symmetric collision system, pp, whereas only azimuthal flattening is used for p--Pb collisions. 

Corrections for tracking inefficiencies are obtained from Monte Carlo simulations with different event generators and particle transport through the detector performed with GEANT3~\cite{Brun:1082634} including a detailed description of the detector conditions during the 2010 and 2013 data taking periods. 
Simulations with the PYTHIA~6 event generator~\cite{Sjostrand:2006za} (Perugia 2011 tune~\cite{Skands:2010ak}) are used to determine the track reconstruction  efficiency for the data sample of  pp collisions.  For the p--Pb system, the  DPMJET event generator~\cite{Roesler:2000he} is used. The $p_{\rm T}$ and $\eta$  dependence of  the single particle detection efficiency is computed based on the ratio of the number of reconstructed tracks from the simulation (known as detector level), corrected for the non-uniformity of the detector (weights),  to the number of generated particles  (known as generator level) as a function of those two variables. Reconstructed tracks from the data sample are corrected for detector non-uniformity and for tracking inefficiencies for extracting the described  corrected correlators.
The number of fully corrected measured charged tracks is averaged over the number of events to extract the quoted $[N_{\rm ch}]$ per multiplicity class.

\section{Statistical and systematic uncertainties}
\label{sec:uncertainties}

The statistical uncertainties on the strengths of the $G_{2}$ correlators are extracted with the sub-sampling method using ten sub-samples for both systems, pp and p--Pb, whereas systematic uncertainties are assessed by repeating the analysis with  different event and track selection criteria. 
The significance of the deviations  with respect to the default analysis conditions is assessed according to a statistical test~\cite{Barlow:2002yb}. The total systematic uncertainties are computed as  quadratic sums of the significant systematic deviations. The contributions to the uncertainty due to the event selection and the kinematic acceptance are estimated by narrowing to 3~cm and expanding to 10~cm the selected range for the distance of the PV to the nominal interaction point along the beam direction. 
Possible biases associated with contamination by secondary particles  are estimated by using  track selection criteria that only require information from the TPC and  relaxing the accepted DCA range. The possible biases in the determination of the track parameters for tracks crossing the TPC in the azimuthal regions close to the sector boundaries is estimated by excluding tracks that lie within those sections from the analysis. This additional selection criterion eliminates  distortions possibly encountered near sector boundaries but produces a nominal 25\% reduction of the azimuthal acceptance.  Track losses are however compensated for by the robust nature of the $G_2$ correlator definition as a ratio of two-particle density to the product of single-particle densities. 
The overall accuracy  of the analysis procedure is additionally estimated by means of a MC closure test. Deviations from  perfect closure are conservatively added to  systematic uncertainties when significant. The same criteria are followed to extract the statistical and systematic uncertainties on $[N_{\rm ch}]$.

As in the   study  of $G_2$ in Pb--Pb collisions reported in Ref.~\cite{ALICE:2019smr}, measurements of the $G_{2}$ longitudinal and azimuthal projections in pp and p--Pb collisions feature  an overall amplitude  uncertainty. This uncertainty includes  correlated (i.e.~common to all bins) and uncorrelated bin-by-bin contributions. The correlated contribution is the average  deviation along all bins while the uncorrelated contribution is what  remains after subtracting such average from  the actual deviation on a per bin basis. 
The largest contribution to the correlated systematic  uncertainties arises from the variation of the track selection criteria with an average value in the different multiplicity classes of 10\% (4\%) for both  the longitudinal and the azimuthal projections  of the $G_{2}^{\rm CD}$ ($G_{2}^{\rm CI}$) correlator in the pp system, and about 12\% (1.5\%) for both the longitudinal and the azimuthal projections in p--Pb collisions, while the other checks have  negligible contributions.
The largest systematic contribution to the uncorrelated uncertainty also stems  from track selection criteria tests with  average values of 6\% (1.5\%) for both the longitudinal and the azimuthal projections of the $G_{2}^{\rm CD}$ ($G_{2}^{\rm CI}$) correlator in the pp system, and less than 9\% (1\%) in the p--Pb system. Total average uncorrelated systematic uncertainties values are approximately the same for the azimuthal and longitudinal projections except for the azimuthal projections of the $G_{2}^{\rm CD}$ correlator in the p--Pb system which reach a 12\% due to the impact of the TPC sector boundaries.

\section{Results}
\label{sec:results}
\begin{figure}[ht]
  \centering
  \includegraphics[scale=0.3,keepaspectratio=true,clip=true,trim=2pt 4pt 43pt 2pt]
  {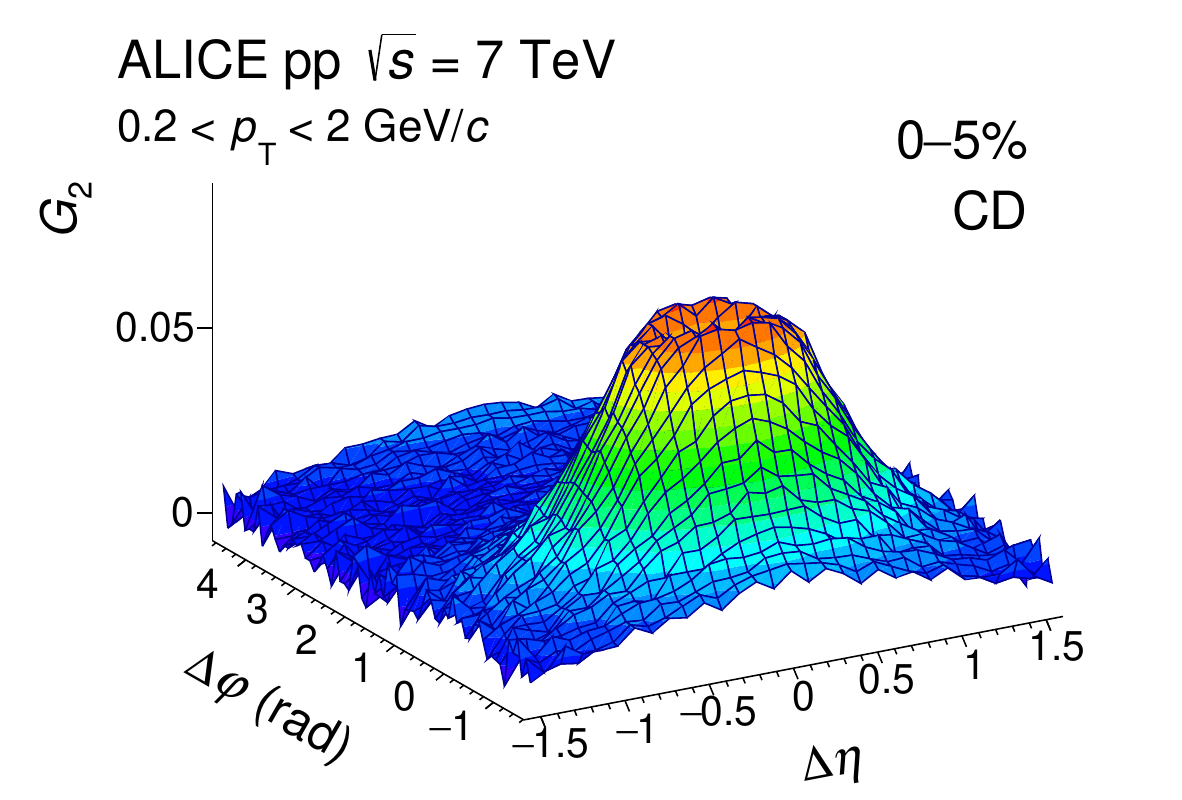}
  \includegraphics[scale=0.3,keepaspectratio=true,clip=true,trim=45pt 4pt 43pt 2pt]
  {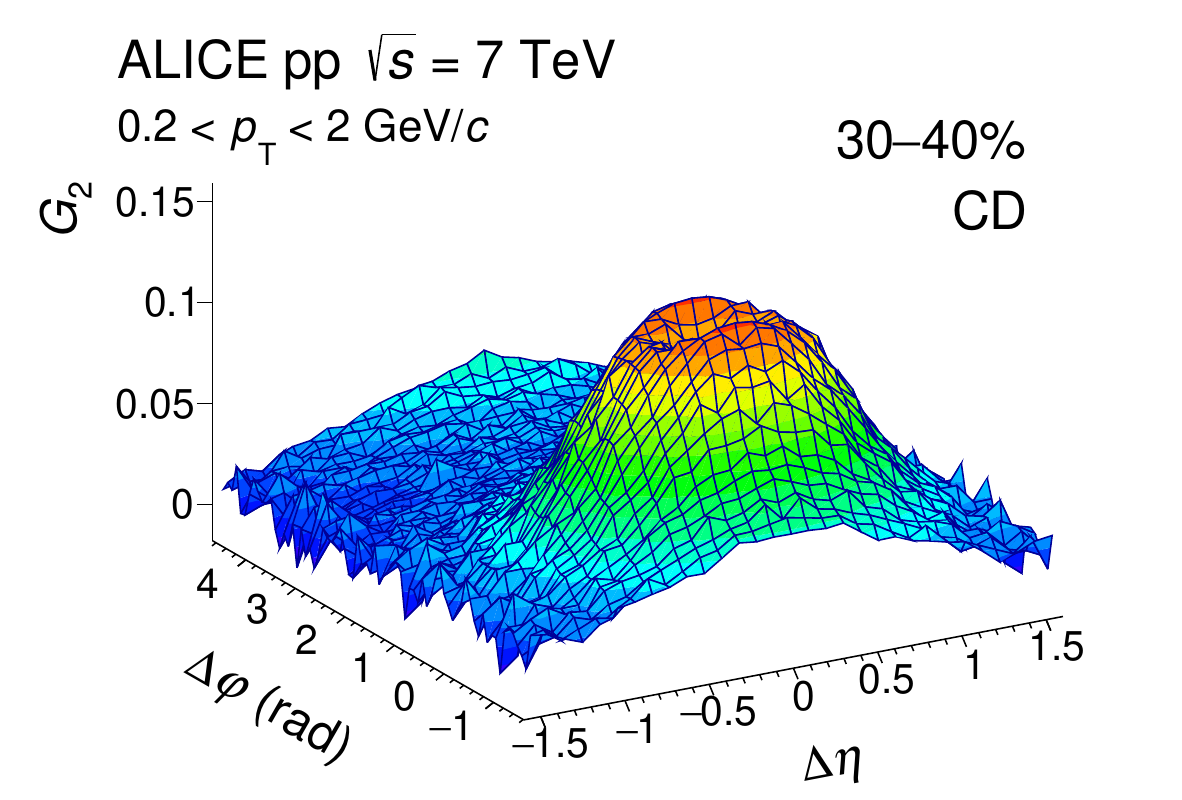}
  \includegraphics[scale=0.3,keepaspectratio=true,clip=true,trim=45pt 4pt 43pt 2pt]
  {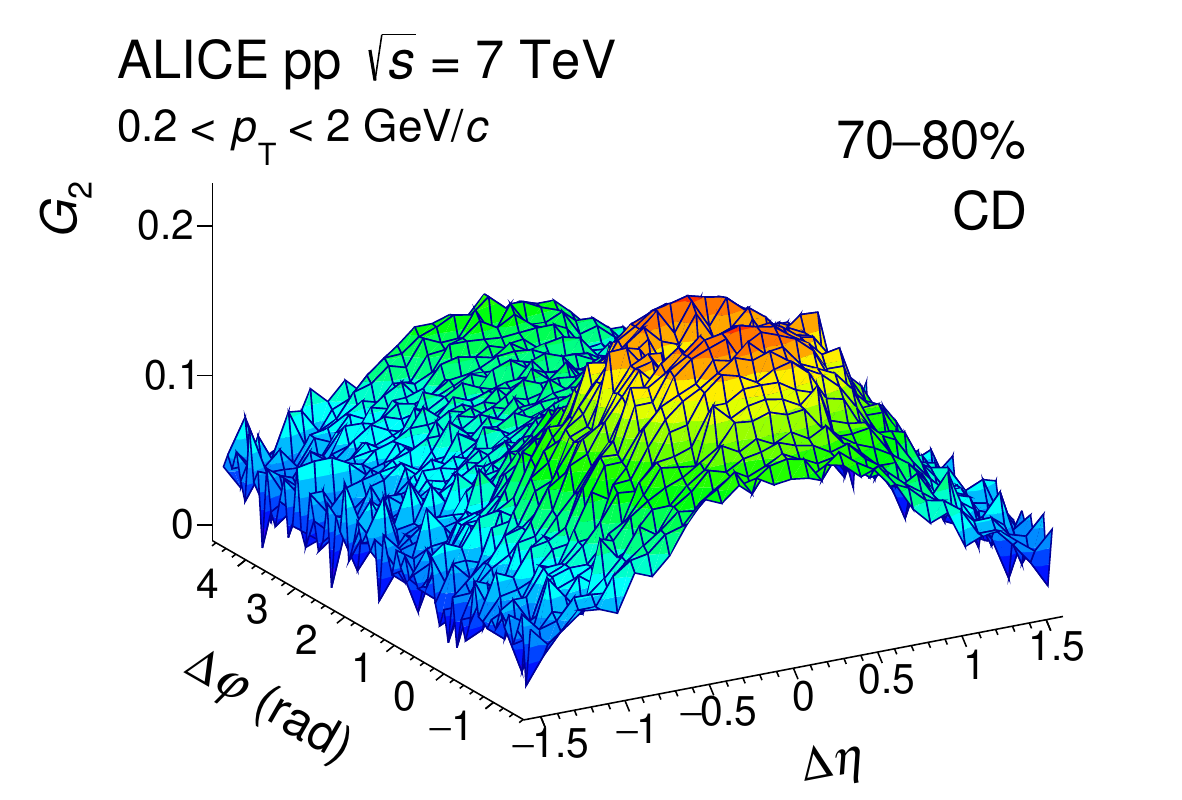} \par
  \includegraphics[scale=0.3,keepaspectratio=true,clip=true,trim=2pt 4pt 43pt 2pt]
  {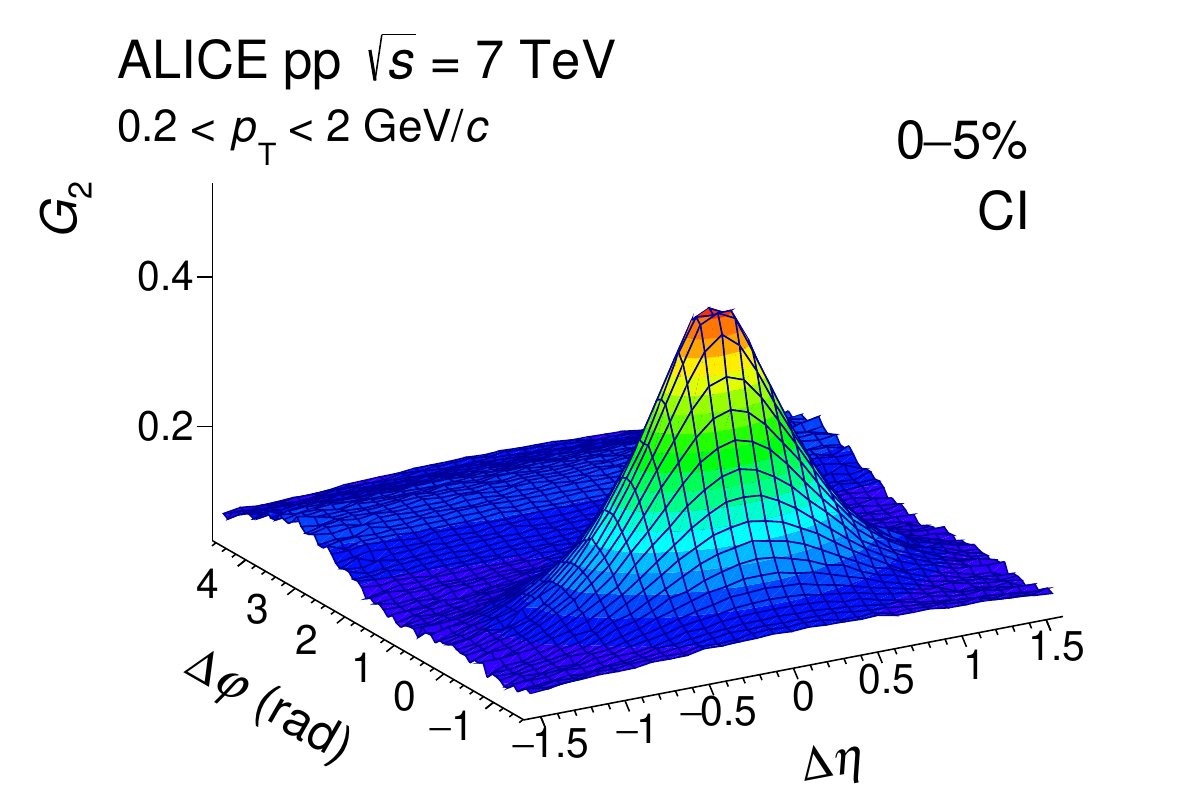}
  \includegraphics[scale=0.3,keepaspectratio=true,clip=true,trim=45pt 4pt 43pt 2pt]
  {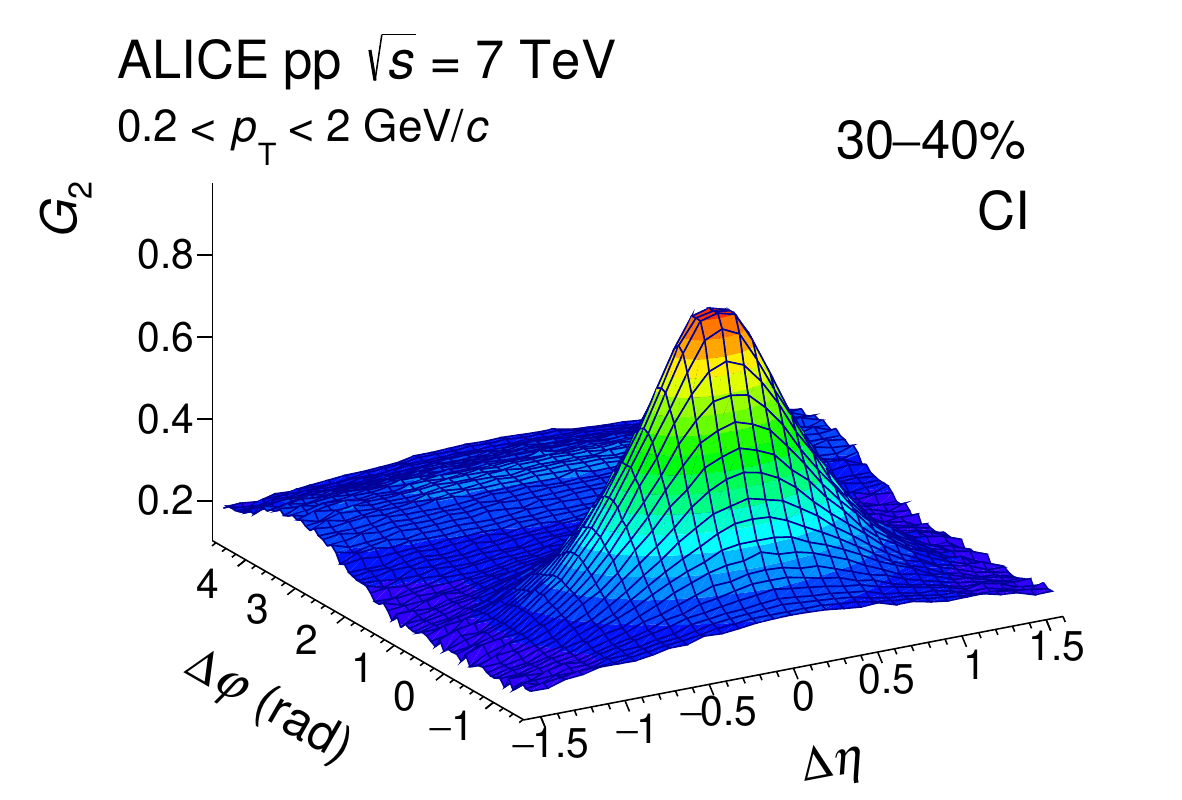}
  \includegraphics[scale=0.3,keepaspectratio=true,clip=true,trim=45pt 4pt 43pt 2pt]
  {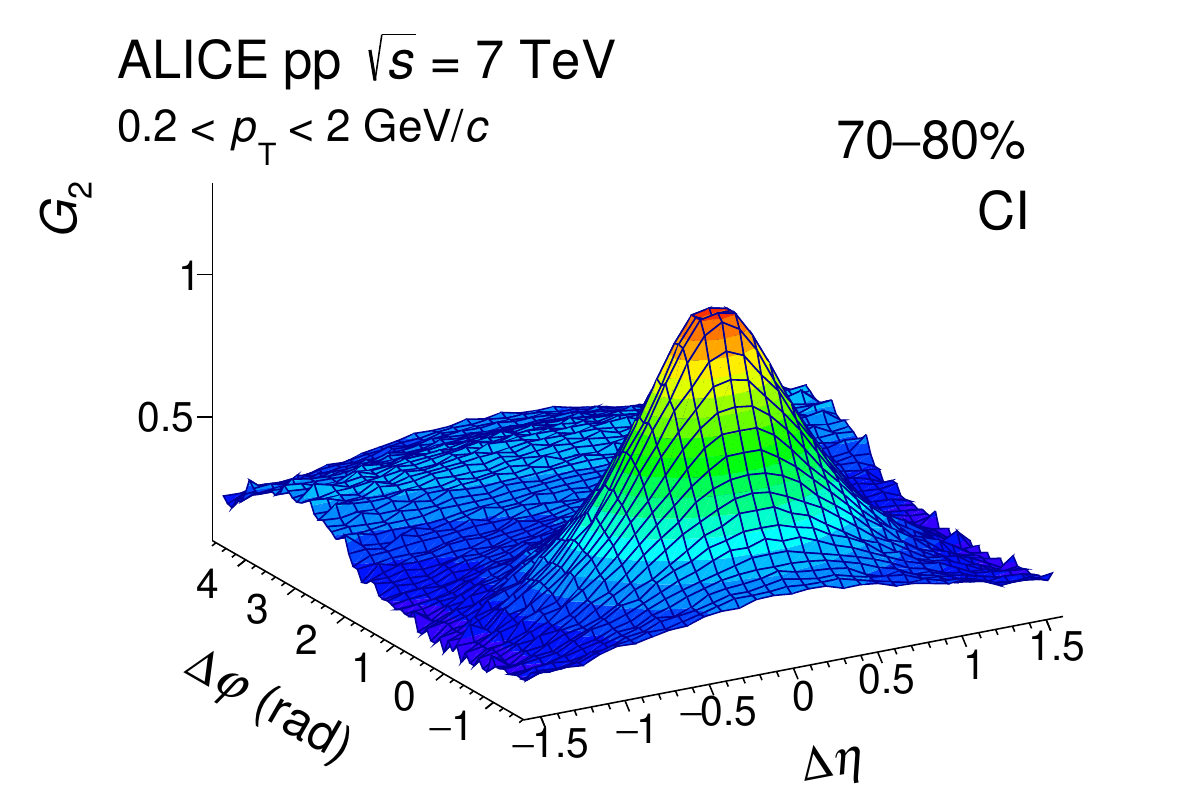} \par
  \caption{\label{fig:2dg2cicdpp} Two-particle transverse momentum correlations $G_{2}^{\rm CD}$ (top) and  $G_{2}^{\rm CI}$ (bottom) for the largest (left), medium (centre) and lowest (right) charged particle multiplicity classes in pp collisions at $\sqrt{s}=7\;\text{TeV}$. The correlator values are not shown in the intervals $|\Delta\eta|<0.1$ and $|\Delta\varphi|<0.09$, which are affected by track merging effects (see text for details).}
\end{figure}

\begin{figure}[hb]
  \centering
  \includegraphics[scale=0.3,keepaspectratio=true,clip=true,trim=2pt 4pt 43pt 2pt]
  {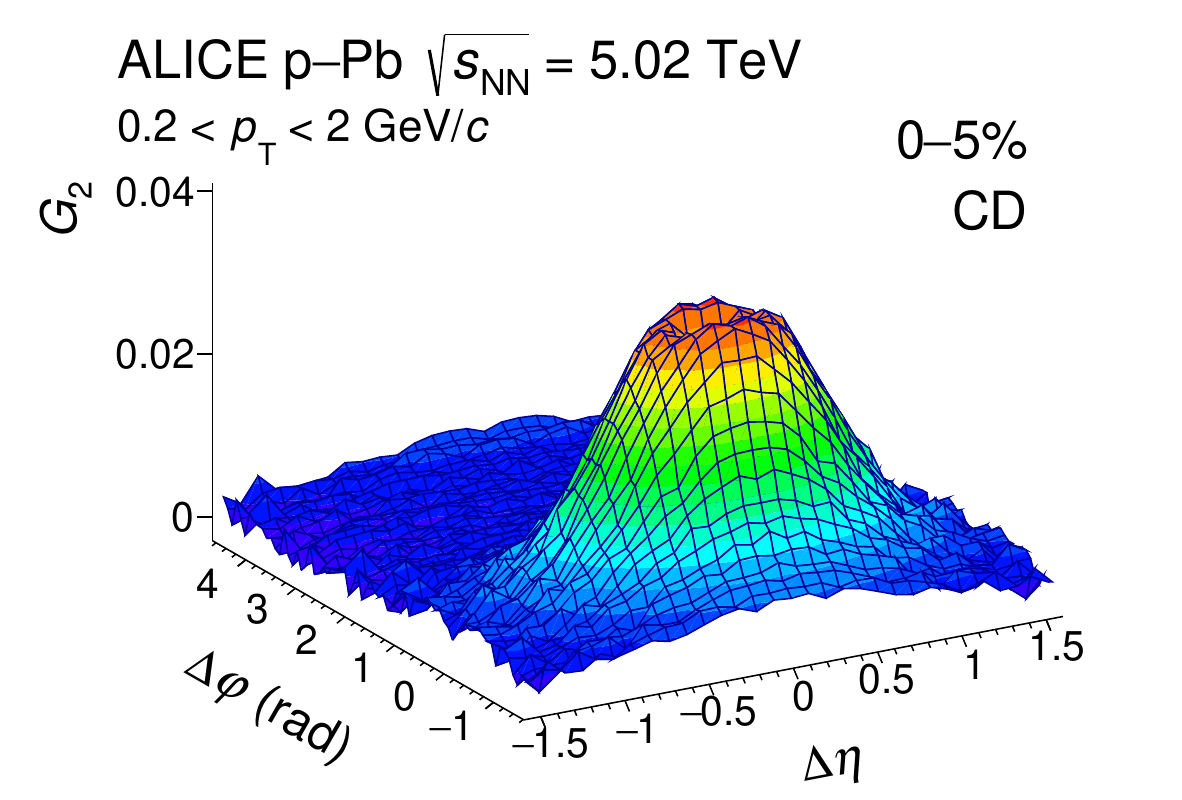}
  \includegraphics[scale=0.3,keepaspectratio=true,clip=true,trim=45pt 4pt 43pt 2pt]
  {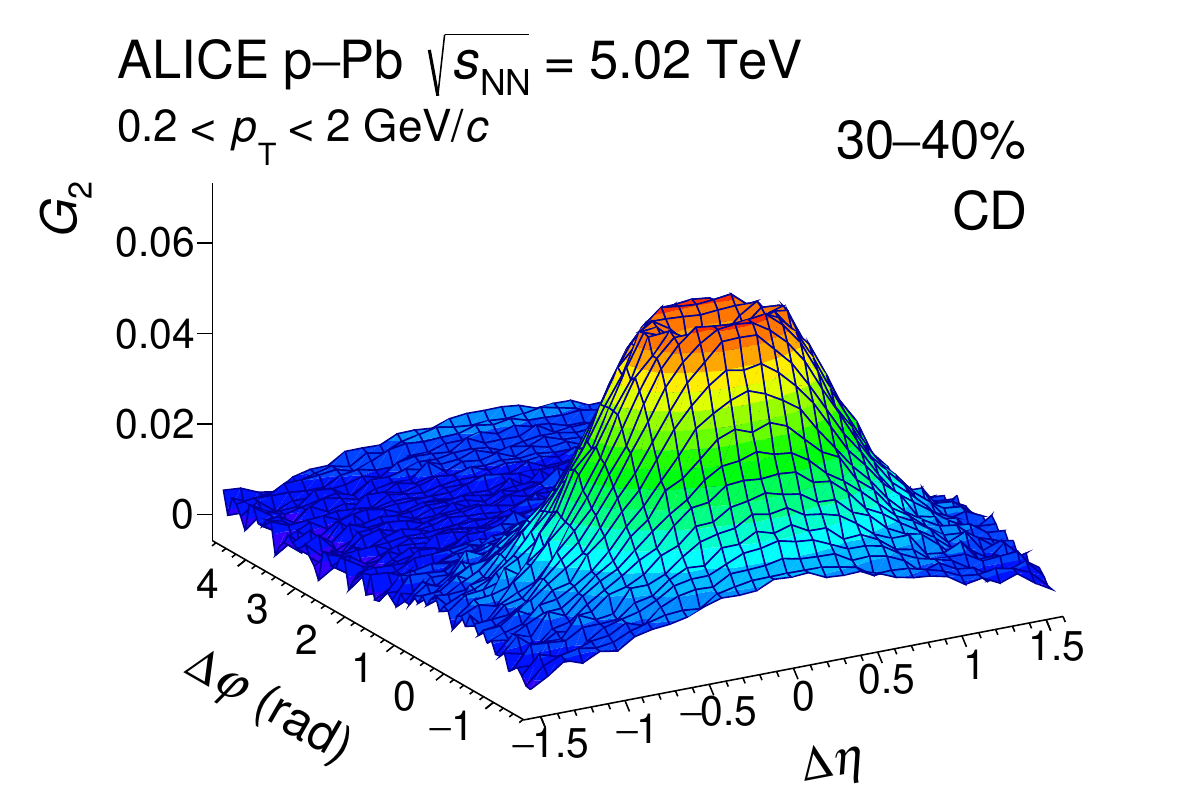}
  \includegraphics[scale=0.3,keepaspectratio=true,clip=true,trim=45pt 4pt 43pt 2pt]
  {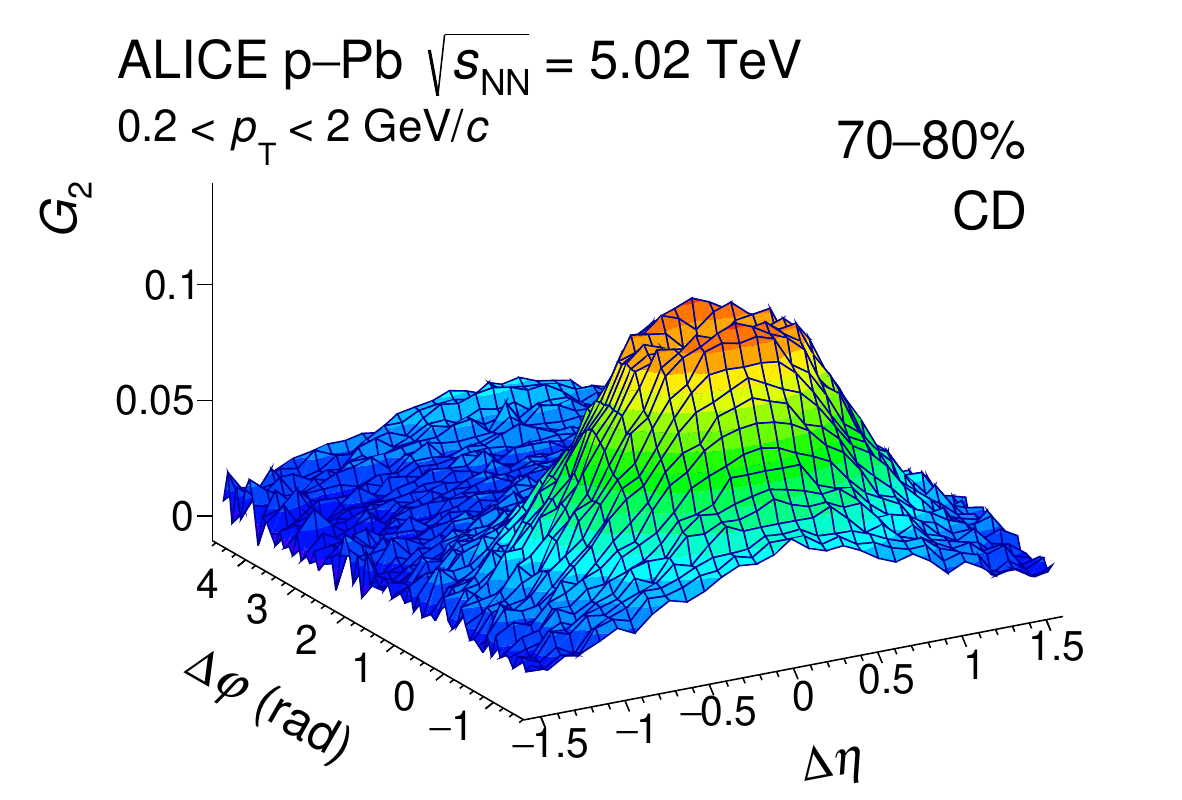} \par
  \includegraphics[scale=0.3,keepaspectratio=true,clip=true,trim=2pt 4pt 43pt 2pt]
  {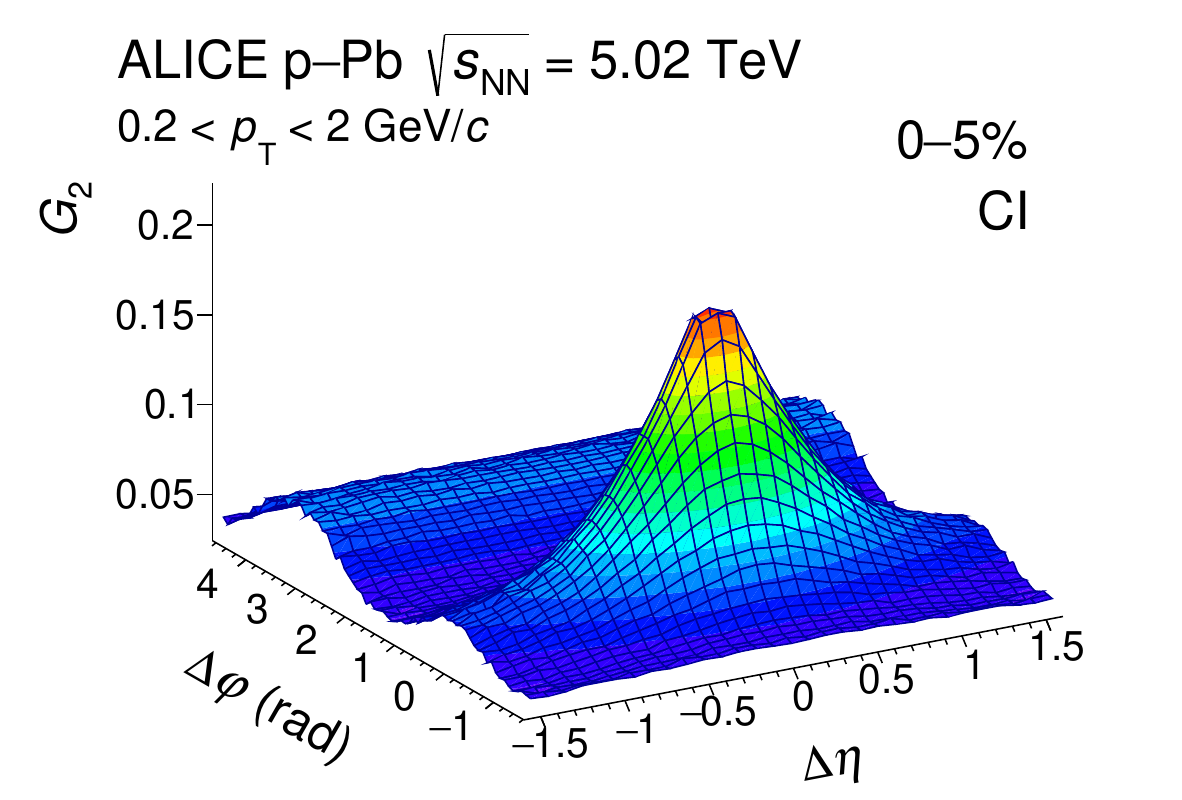}
  \includegraphics[scale=0.3,keepaspectratio=true,clip=true,trim=45pt 4pt 43pt 2pt]
  {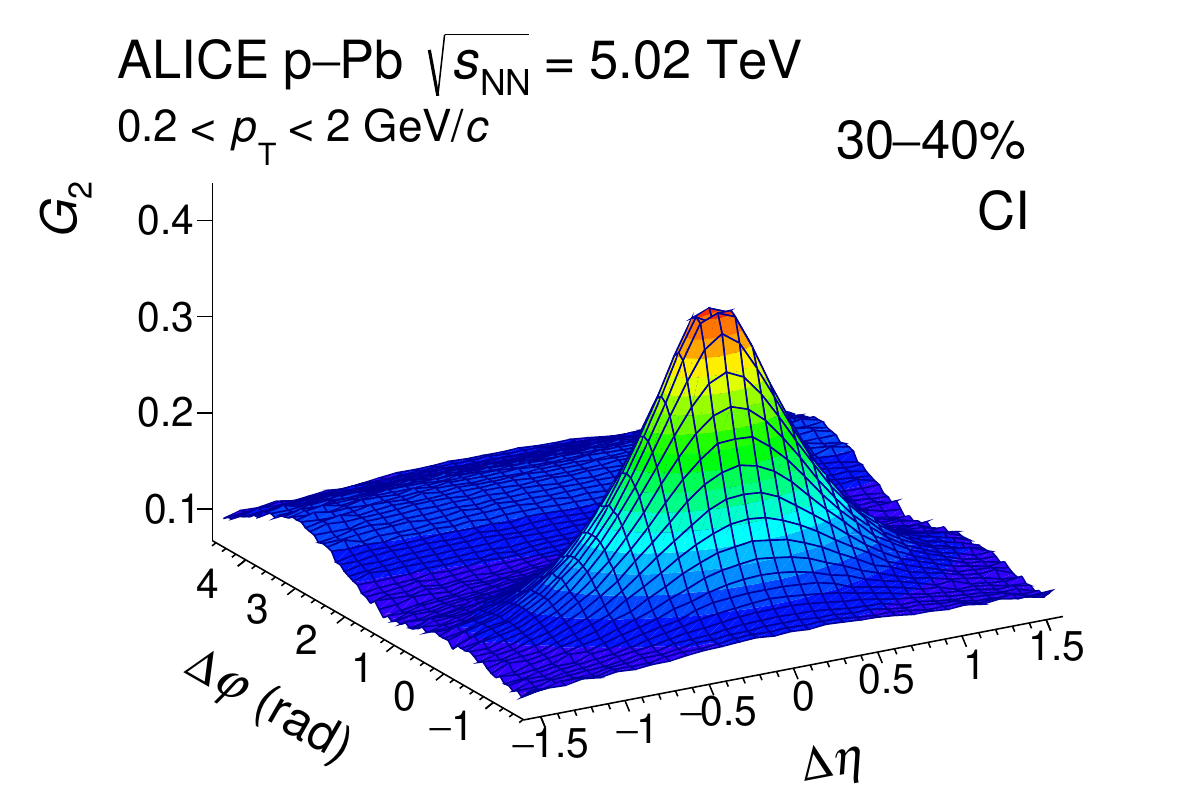}
  \includegraphics[scale=0.3,keepaspectratio=true,clip=true,trim=45pt 4pt 43pt 2pt]
  {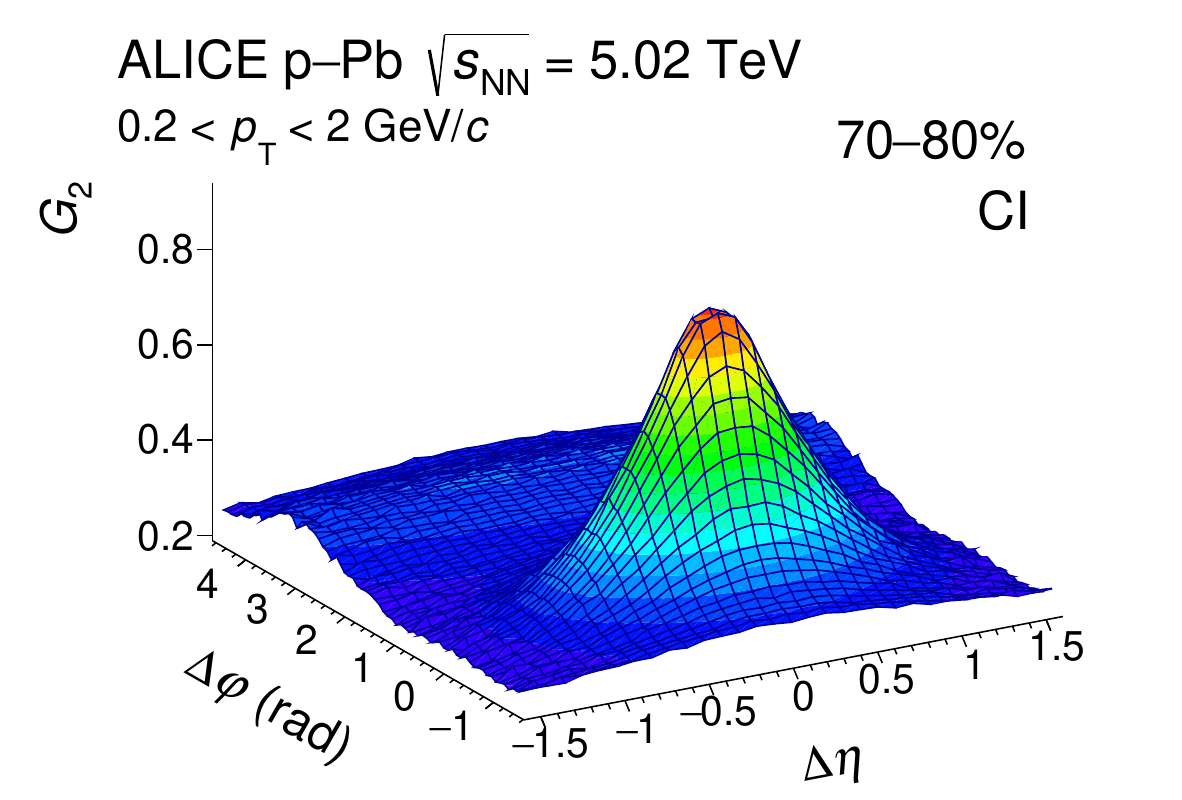} \par
  \caption{\label{fig:2dg2cicdppb} Two-particle transverse momentum correlations $G_{2}^{\rm CD}$ (top) and $G_{2}^{\rm CI}$ (bottom) for the largest (left), medium (centre), and lowest (right) charged particle multiplicity classes in p--Pb collisions at $\sqrt{s_{\rm NN}}=5.02\;\text{TeV}$. The correlator values are not shown in the intervals $|\Delta\eta|<0.1$ and $|\Delta\varphi|<0.09$, which are affected by track merging effects (see text for details).}
\end{figure}

The $G_{2}^{\rm CD}$ and $G_{2}^{\rm CI}$ correlators measured in pp collisions at $\sqrt{s} = 7\;\text{TeV}$ and p--Pb collisions at $\sqrt{s_{\rm NN}} = 5.02\;\text{TeV}$ are shown in Figs.~\ref{fig:2dg2cicdpp} and~\ref{fig:2dg2cicdppb}, respectively, for three selected multiplicity classes,
as functions of the pair separation in pseudorapidity $\Delta\eta$ and azimuth $\Delta\varphi$.
The  $G_{2}^{\rm CD}$ and $G_{2}^{\rm CI}$ correlators  exhibit common features in both pp and p--Pb collisions. Such features include  a prominent  peak centered at $\Delta\eta=0$, $\Delta\varphi=0$, hereafter referred to as the near-side peak, and a relatively flat plateau shaped distribution   surrounding $\Delta\varphi=\pi$, known as the away side,  and extending across the $\Delta\eta$ acceptance of the measurement.
The near-side peak of  both $G_{2}^{\rm CD}$ and $G_{2}^{\rm CI}$  exhibits a  monotonically decreasing amplitude   from the  lowest to the highest multiplicity  classes, in both collision systems, while the peak  shapes  are approximately independent of the collision  multiplicity. 
It is also observed that the away-side amplitude of the $G_{2}^{\rm CD}$ correlator measured in pp collisions decreases somewhat faster than  that of the near-side peak, whereas the shape of the away side of the $G_{2}^{\rm CI}$ correlators exhibits only modest variations with multiplicity. 
 
Additionally, the
$G_{2}^{\rm CI}$  correlators measured in p--Pb collisions   feature a modest azimuthal modulation approximately uniform in magnitude across the $\Delta\eta$ range of the measurement. The modulation is expected from prior ALICE azimuthal correlation measurements~\cite{ALICE:2012eyl} but  observed in greater detail in Fig.~\ref{fig:2dg2cicdppb}. 
Remarkably, also  the near-side peak of the CD correlator in p--Pb collisions does not appear to narrow significantly with increasing  multiplicity in contrast with the behaviour observed in  the pp system.

\begin{figure}[ht]
  \centering
  \includegraphics[width=0.45\textwidth,keepaspectratio=true,clip=true,trim=2pt 4pt 43pt 2pt]
  {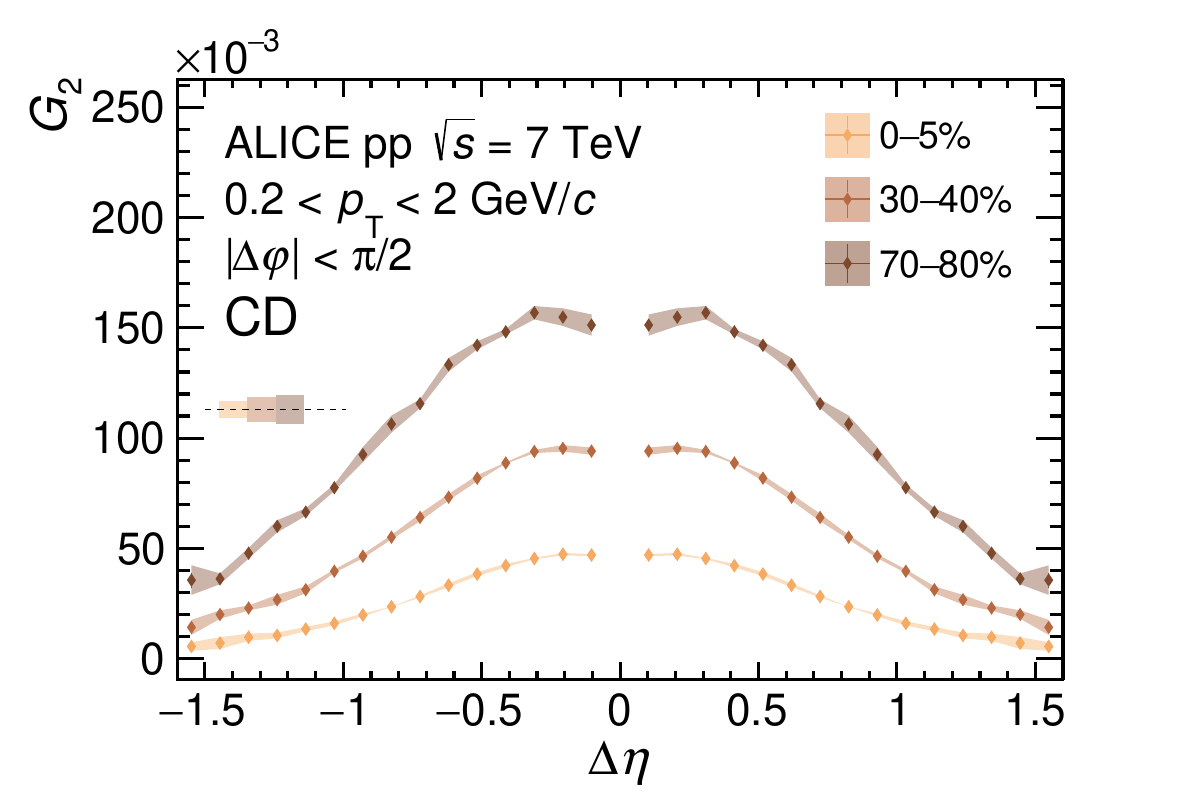}
  \includegraphics[width=0.45\textwidth,keepaspectratio=true,clip=true,trim=2pt 4pt 43pt 2pt]
  {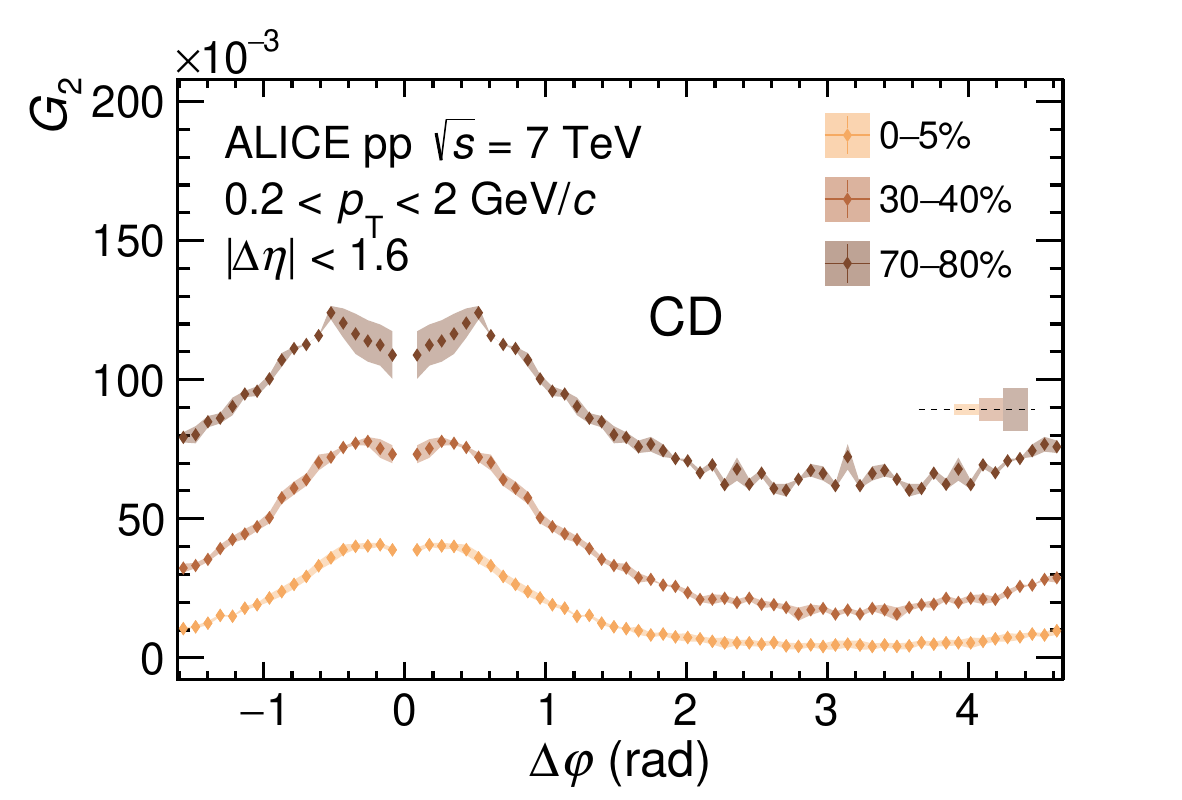} \par
  \includegraphics[width=0.45\textwidth,keepaspectratio=true,clip=true,trim=2pt 4pt 43pt 2pt]
  {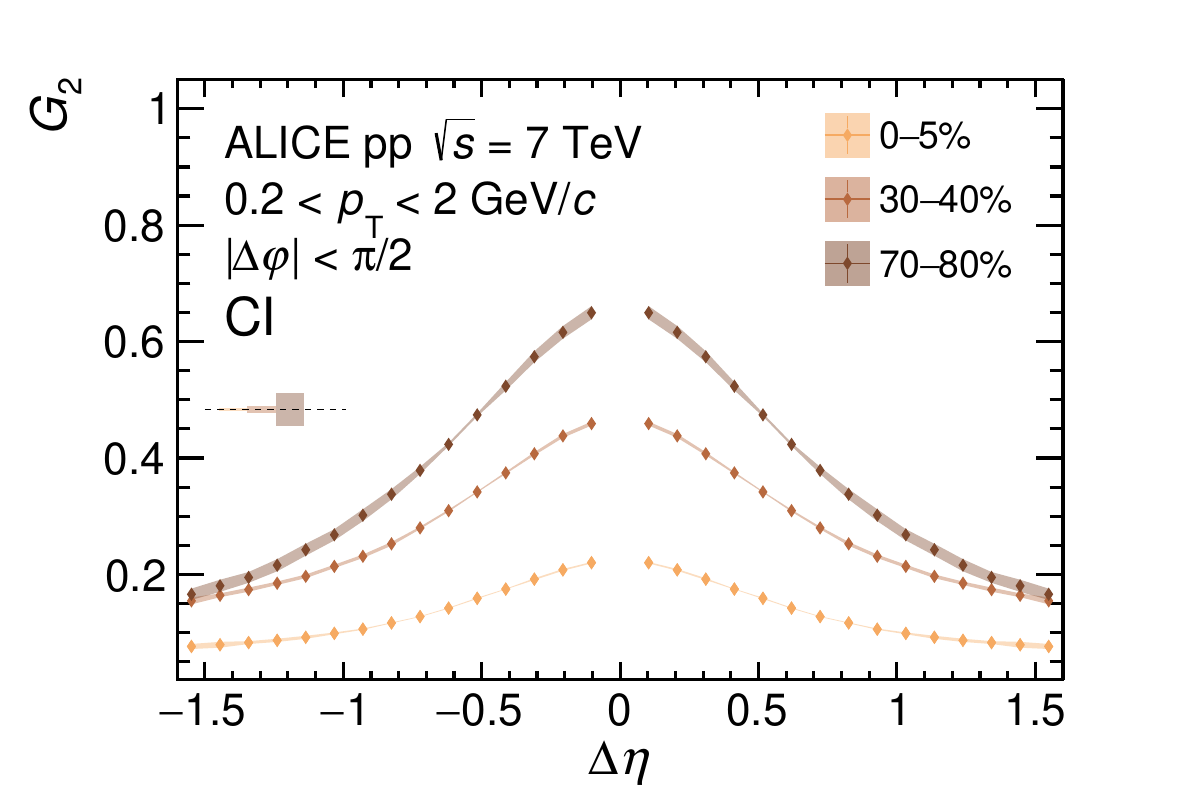}
  \includegraphics[width=0.45\textwidth,keepaspectratio=true,clip=true,trim=2pt 4pt 43pt 2pt]
  {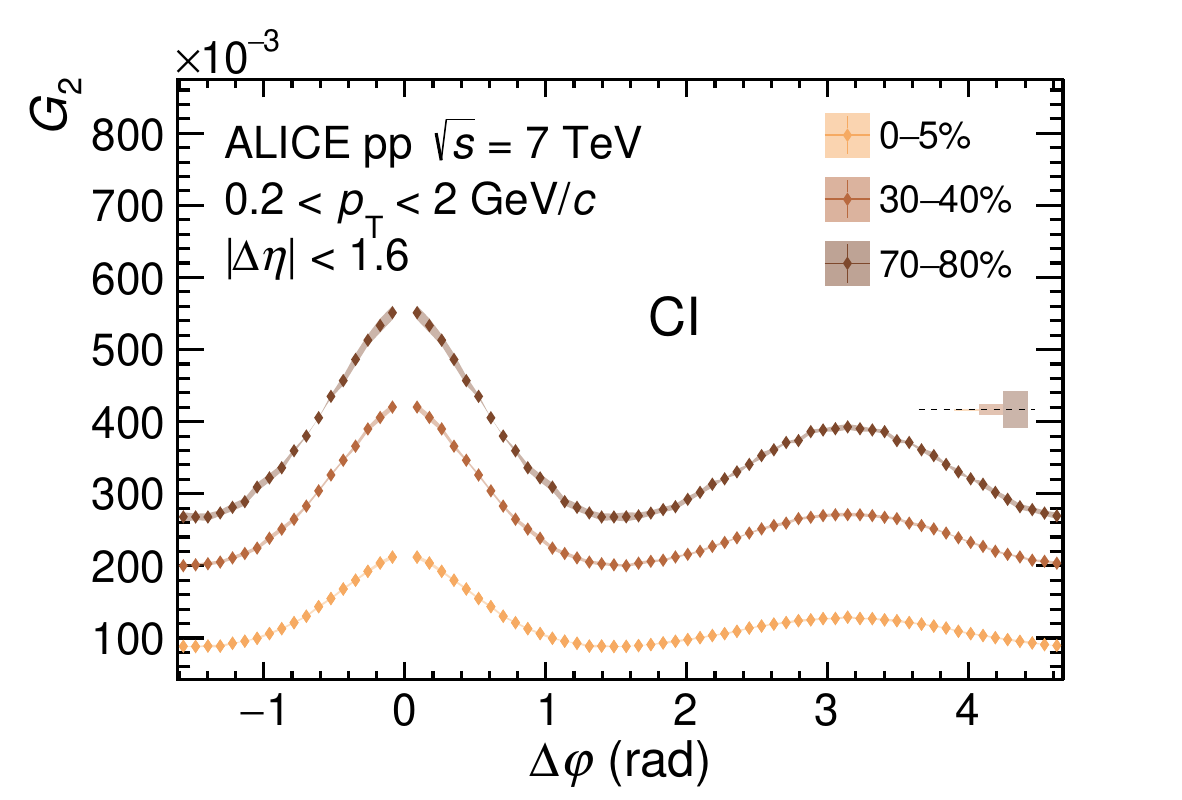} \par
  \caption{\label{fig:detadphig2cicdpp} Longitudinal (left) and azimuthal (right) projections of the two-particle transverse momentum correlations $G_{2}^{\rm CD}$ (top) and $G_{2}^{\rm CI}$ (bottom) for selected charged particle multiplicity classes in pp collisions at $\sqrt{s}=7\;\text{TeV}$. The correlator values are not shown in the intervals $|\Delta\eta|<0.1$ and $|\Delta\varphi|<0.09$, which are affected by track merging effects (see text for details). Vertical bars (mostly smaller than the marker size) and shaded bands represent statistical and uncorrelated systematic uncertainties, respectively. Correlated systematic uncertainties are represented as small boxes at the sides of the panels.}
\end{figure}

\begin{figure}[ht]
  \centering
  \includegraphics[width=0.45\textwidth,keepaspectratio=true,clip=true,trim=2pt 4pt 43pt 2pt]
  {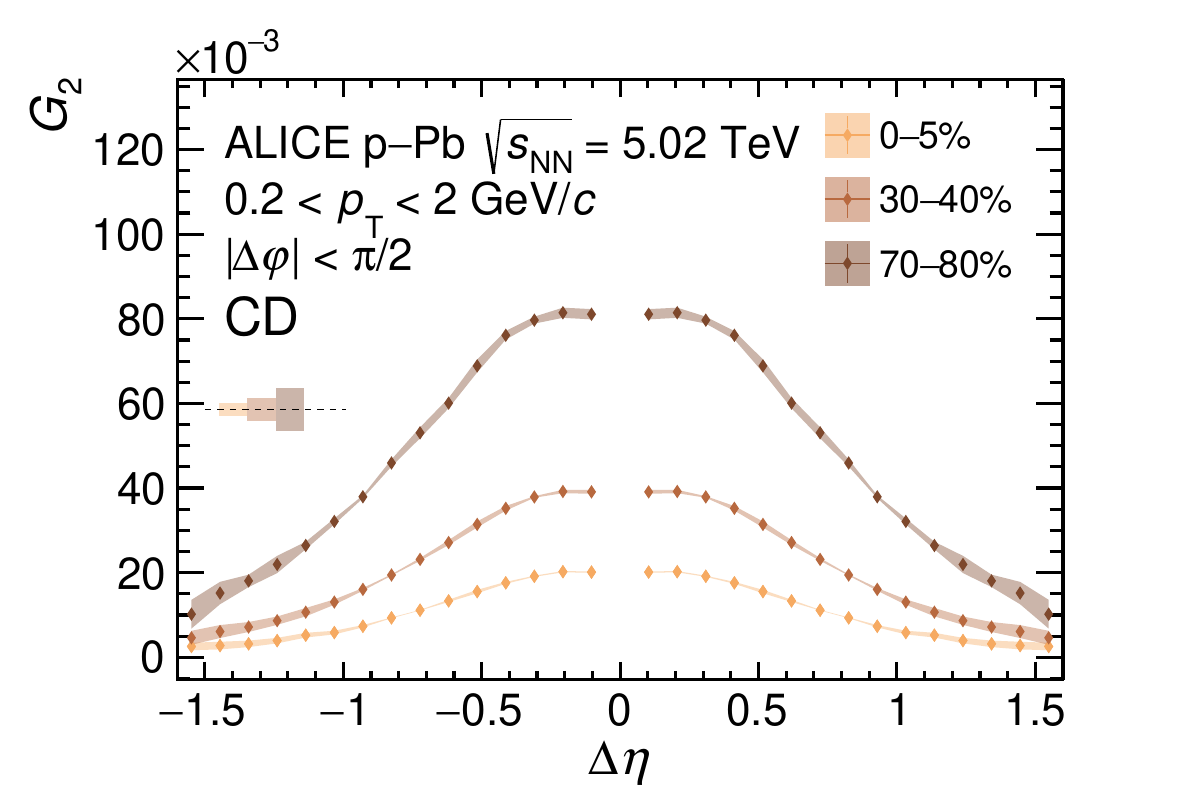}
  \includegraphics[width=0.45\textwidth,keepaspectratio=true,clip=true,trim=2pt 4pt 43pt 2pt]
  {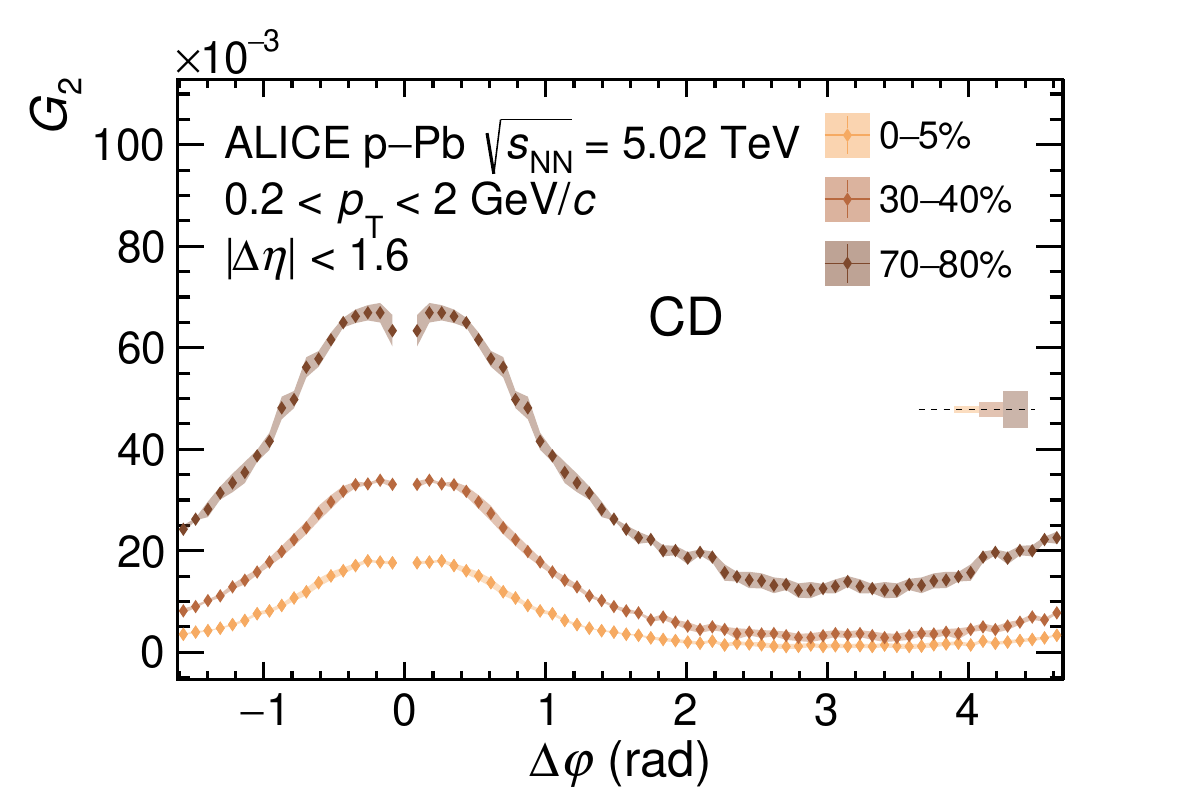} \par
  \includegraphics[width=0.45\textwidth,keepaspectratio=true,clip=true,trim=2pt 4pt 43pt 2pt]
  {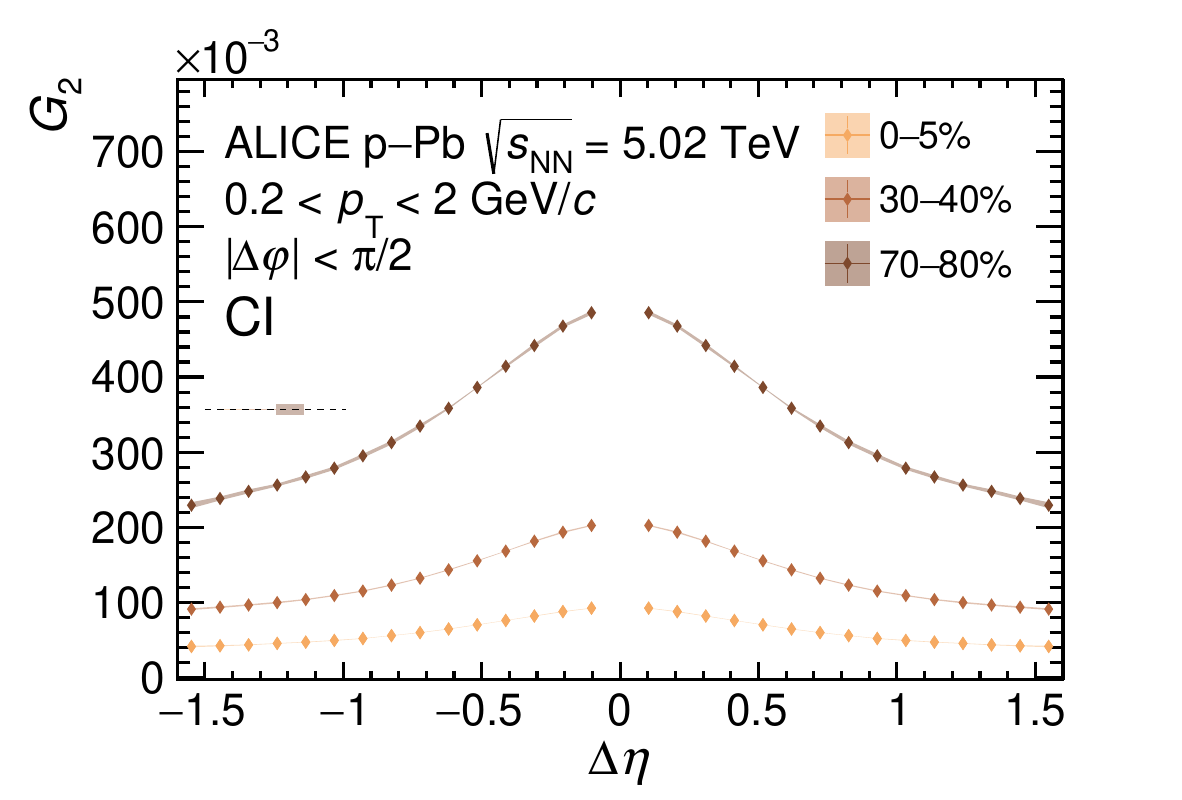}
  \includegraphics[width=0.45\textwidth,keepaspectratio=true,clip=true,trim=2pt 4pt 43pt 2pt]
  {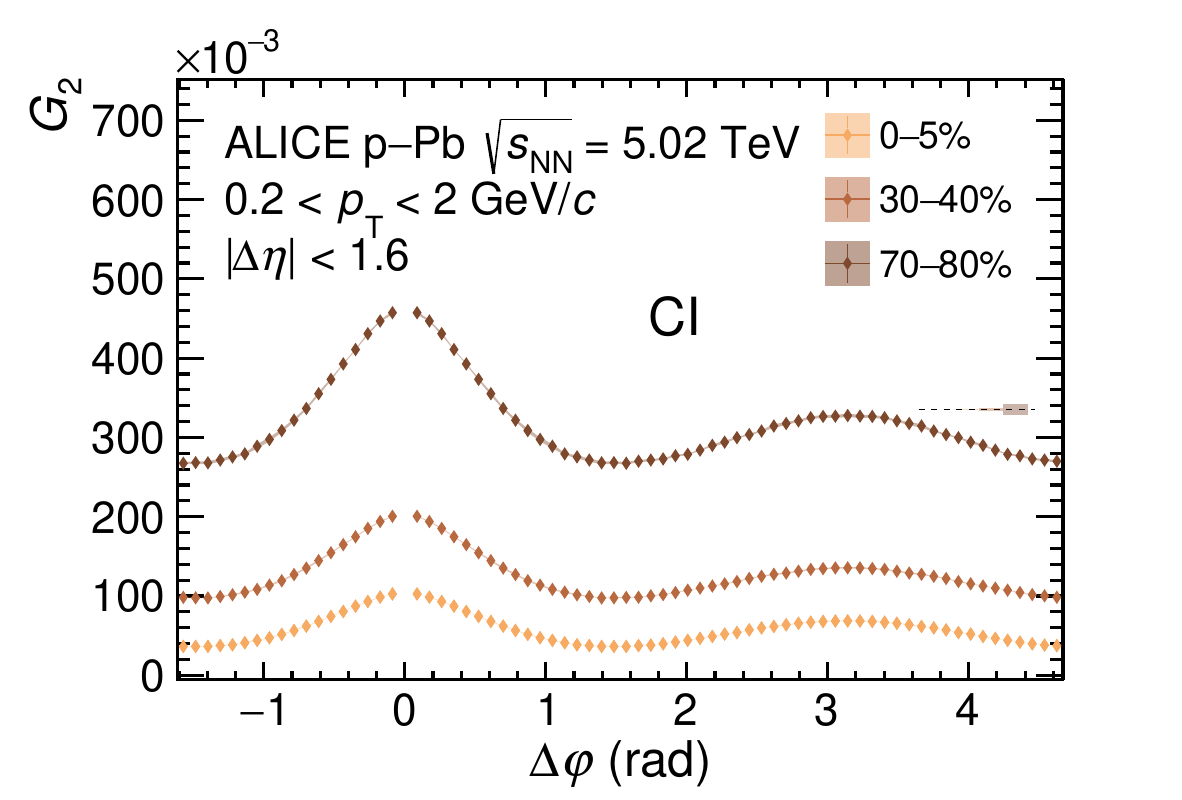}\par
  \caption{\label{fig:detadphig2cicdppb} Longitudinal (left) and azimuthal (right) projections of the two-particle transverse momentum correlations $G_{2}^{\rm CD}$ (top) and $G_{2}^{\rm CI}$ (bottom) for selected charged particle multiplicity classes in p--Pb collisions at $\sqrt{s_{\rm NN}}=5.02\;\text{TeV}$. The correlator values are not shown in the intervals $|\Delta\eta|<0.1$ and $|\Delta\varphi|<0.09$, which are affected by track merging effects (see text for details). Vertical bars (mostly smaller than the marker size) and shaded bands represent statistical and uncorrelated systematic uncertainties, respectively. Correlated systematic uncertainties are represented as small boxes at the sides of the panels.}
\end{figure}

Further examination of the  evolution of the correlators as a function of the produced particle multiplicity is done  by studying their longitudinal and azimuthal  projections.  The longitudinal projections, shown in the left panels of Figs.~\ref{fig:detadphig2cicdpp} and~\ref{fig:detadphig2cicdppb} for the pp and p--Pb systems, respectively, display the average of the $G_{2}$ correlators as a function of $\Delta\eta$ for the near-side azimuthal interval $|\Delta\varphi|<\pi/2$ whereas the azimuthal projections (right panels) are obtained by averaging the correlators over the full $\Delta\eta$ range of the measurements. The ranges $|\Delta\eta|<0.1$ and $|\Delta\varphi|<0.09$  are subject to track merging effects difficult to properly correct for and are thus omitted in the projection plots. The longitudinal projections of the CI correlators (bottom panels) obtained in both pp and p--Pb  collisions feature broad Gaussian-like peaks versus $\Delta\eta$ whose magnitude decreases with increasing multiplicity. The azimuthal CI projections, by contrast, feature a modulation yielding two maxima: the first, centered at $\Delta\varphi=0$, corresponds to the near-side peak of the correlation functions displayed in Figs.~\ref{fig:2dg2cicdpp} and \ref{fig:2dg2cicdppb}, whereas the second maximum reflects the broad away-side peak of these functions. 

The projections of the CD correlators, presented in Figs.~\ref{fig:detadphig2cicdpp} and~\ref{fig:detadphig2cicdppb} (top panels), feature a somewhat more complex dependence on $\Delta \eta$ and $\Delta \varphi$ than those of the CI correlators. In particular, in contrast to the Gaussian-like peaks seen in the CI projections, the CD longitudinal projections exhibit small and narrow dips atop the peak, which is broader than that of the CI projections. 
The dip is most prominent in the azimuthal projections of the $G_2^{\rm CD}$ correlator and largest in the 70--80\% multiplicity class of   pp collisions.   The width and depth of the dip are manifestly  functions of the multiplicity of the collisions and thus appear to feature a system size dependence. 

This system size dependence  is qualitatively understood as arising, in part, from femtoscopic (HBT) correlations.  The CD correlators are computed as the difference between US and LS correlators. The $G_2^{\rm LS}$ correlators, much like the femtoscopic (number) correlation functions measured as a function of the invariant momentum difference of particle pairs, are sensitive to the presence of bosonic interference,  and it is well established that the widths of these correlation functions are inversely proportional to the size of the measured system~\cite{Jeon:2001ue}. It is thus expected that $G_2^{\rm LS}$,  measured as function of $\Delta\eta$ or $\Delta\varphi$, should also exhibit such a  dependence on the system size.  This dependence is seen as dips because the LS correlators are subtracted from the US correlators. Additionally, note that the effect is smaller in p--Pb collisions, most likely because of the larger size of the systems formed in these collisions. 

\section{Shape evolution with multiplicity}
\label{sec:shapeextraction}

The multiplicity evolution of the shape and strength of 
the $G_2^{\rm{CD}}$ and $G_2^{\rm{CI}}$ correlators measured in pp and p--Pb collisions
is analysed using a multicomponent model already utilised in Pb--Pb collisions~\cite{ALICE:2019smr} and  defined as
\begin{equation}
          F( \Delta \eta, \Delta \varphi) = B + \displaystyle\sum_{\rm n=2}^{6} a_{\rm n} \times \cos \left({\rm n}\, \Delta \varphi\right) + 
  A \times\frac{\gamma_{\Delta \eta}}{2\,\omega_{\Delta \eta}\,
      \Gamma \left(\frac{1}{\gamma_{\Delta \eta}}\right)}
      \,\rm{e}^{- \left| \frac{\Delta \eta}{\omega_{\Delta \eta}}\right|^{\gamma_{\Delta \eta}}}\times
          \frac{\gamma_{\Delta \varphi}}{2\,\omega_{\Delta \varphi}\,
      \Gamma \left(\frac{1}{\gamma_{\Delta \varphi}}\right)}
      \,\rm{e}^{- \left| \frac{\Delta \varphi}{\omega_{\Delta \varphi}}\right|^{\gamma_{\Delta \varphi}}} \text{,}
\label{eq:fitfunc}
\end{equation}
where $B$ and $a_{\rm n}$ are intended to describe the  long-range mean correlation strength and the possible azimuthal aniso\-tropies, while the bidimensional generalised Gaussian, defined by the parameters $A$, $\omega_{\Delta \eta}$, $\omega_{\Delta \varphi}$, $\gamma_{\Delta \eta}$, and $\gamma_{\Delta \varphi}$, is intended to model the near-side peak. 

The main focus of this paper is specifically on measuring the  evolution of the azimuthal and longitudinal widths of the prominent near-side peak of the $G_2^{\rm{CD}}$ and $G_2^{\rm{CI}}$ correlators, which is quantified in terms of width parameters  $\sigma_{\Delta \eta}$ and $\sigma_{\Delta \varphi}$  computed according to
\begin{equation}
  \sigma_{\Delta \eta (\Delta \varphi)} = \sqrt{\frac{\omega^2_{\Delta \eta (\Delta \varphi)} 
      \Gamma(3/\gamma_{\Delta \eta (\Delta \varphi)})}{\Gamma(1/\gamma_{\Delta \eta (\Delta \varphi)})}}\text{.}
\end{equation}

Bidimensional fits to the measured $G_{2}^{\rm CD}$ and $G_{2}^{\rm CI}$ correlators were carried out with the least-squares method, considering only the statistical uncertainties. The central region around $|\Delta \eta| = 0$ and $|\Delta\varphi| = 0$ was excluded from the fit to avoid biases associated with track merging. The excluded region was enlarged, when appropriate, to cover the narrow dip found in the CD correlation functions. The differences between data points and fit functions were examined in detail and found to be negligible relative to the amplitude of the correlation functions except in some areas of the near-side peak tails and close to the excluded patch around $\Delta\eta, \Delta\varphi = (0,0)$, thereby yielding a full fit $\chi^2/{\rm dof}$ in the range 2 to 9. Fits were repeated using systematic uncertainties of the correlation functions to examine the possibility of biases. Widths obtained with these larger uncertainties were within the systematic uncertainties of the nominal values, reported for fits performed  with statistical uncertainties, and the $\chi^2/{\rm dof}$ values dropped below unity.

Systematic uncertainties on the extracted widths were assessed using the procedure  described in Sec.~\ref{sec:uncertainties}. The largest contributor to these uncertainties is the track selection criteria  with  values of  4\% (2\%) and 2\% (2\%) for the longitudinal and azimuthal widths, respectively, of the $G_{2}^{\rm CD}$ ($G_{2}^{\rm CI}$) correlator in the pp system, and 4\% (2\%) for both widths in the p--Pb system. Total systematic uncertainties on the widths amount to 5\% (2\%) and 3\% (2\%) for the longitudinal and azimuthal widths, respectively, of the $G_{2}^{\rm CD}$ ($G_{2}^{\rm CI}$) correlator in the pp system, and 5\% (3\%) and 4\% (2\%) in the p--Pb system.

\begin{figure}[ht]
  \includegraphics[width=0.99\textwidth,keepaspectratio=true,clip=true,trim=0pt 0pt 0pt 0pt]
  {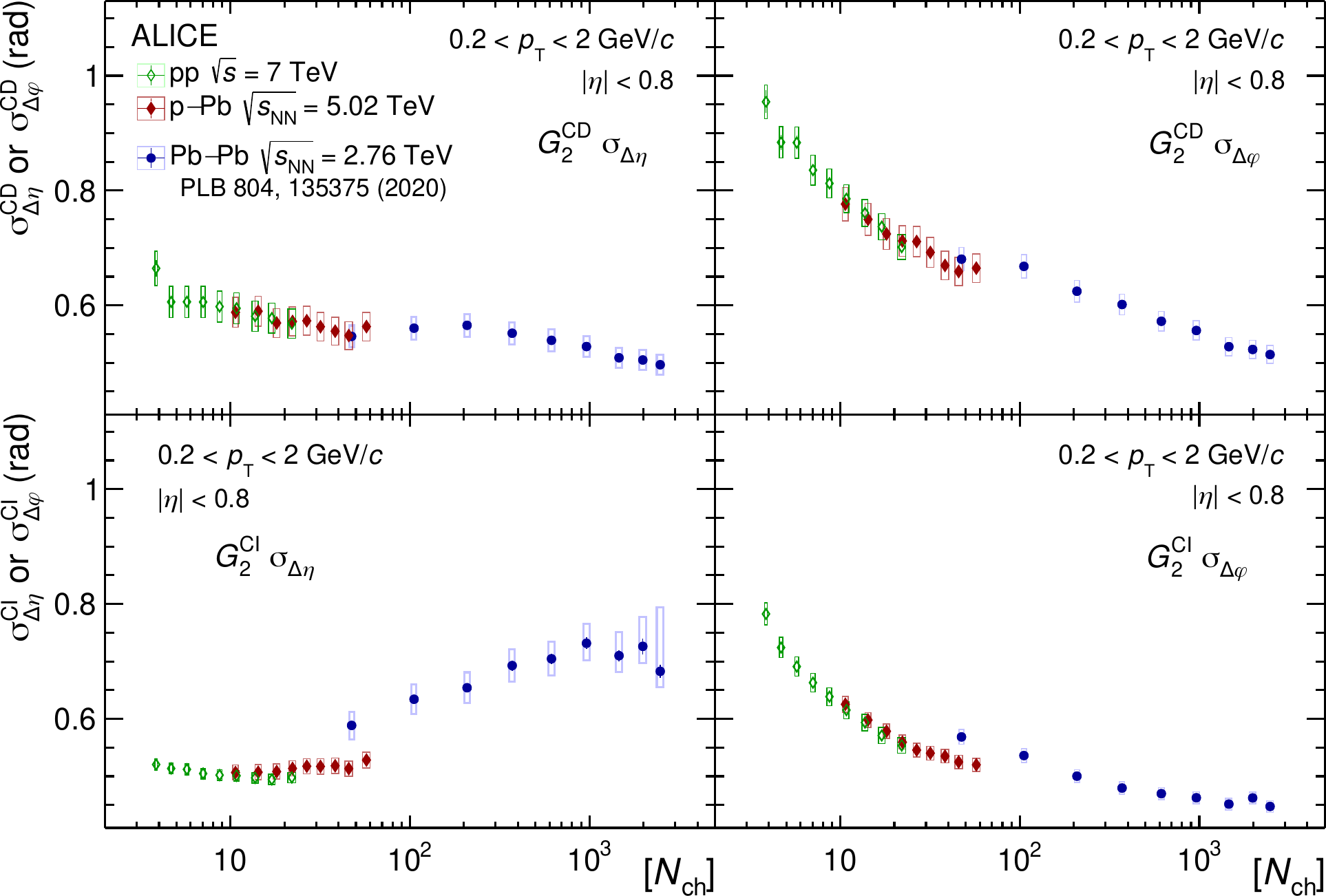}
  \caption{Evolution with the average charged particle multiplicity of the longitudinal (left) and azimuthal (right) widths of the two-particle transverse momentum differential correlation $G_{2}^{\rm CD}$ (top row) and $G_{2}^{\rm CI}$ (bottom row) in pp, p--Pb, and Pb--Pb collisions at $\sqrt{s} = 7\;\text{TeV}$, $\sqrt{s_{\rm NN}} = 5.02\;\text{TeV}$, and $\sqrt{s_{\rm NN}} = 2.76\;\text{TeV}$, respectively. Vertical bars (mostly smaller than the marker size) and filled boxes represent statistical and systematic uncertainties, respectively.}
  \label{fig:ppppbpbpbg2cicd}
\end{figure}

Figure~\ref{fig:ppppbpbpbg2cicd} shows the evolution of the longitudinal and azimuthal widths $\sigma$ of the $G_2^{\rm CD}$ (top panels) and $G_2^{\rm CI}$ (bottom panels) correlators  as a function of the average charged particle multiplicity $[N_{\rm ch}]$. The $G_2^{\rm CD}$ and $G_2^{\rm CI}$ correlators widths measured  in Pb--Pb collisions at $\sqrt{s_{\rm NN}} = 2.76\;\text{TeV}$ by the ALICE Collaboration~\cite{ALICE:2019smr} are also displayed.

First focusing  on the evolution of the azimuthal and longitudinal widths of the $G_2^{\rm CD}$ correlator with charged particle multiplicity shown in the top panels of  Fig.~\ref{fig:ppppbpbpbg2cicd}, it is observed that this correlator exhibits a strong azimuthal narrowing with increasing $[N_{\rm ch}]$ in both pp and p--Pb collisions  and a somewhat weaker  narrowing trend in the longitudinal direction. A qualitatively similar narrowing, first  reported in Ref.~\cite{ALICE:2019smr}, is also observed in Pb--Pb interactions across a broad range of collision multiplicities in both the longitudinal and azimuthal directions. Overall, both the longitudinal and azimuthal widths of the near-side peak in the  $G_2^{\rm CD}$ correlator show a smooth narrowing trend with increasing multiplicity across the three different collision systems here considered. 

Shifting the focus to the bottom panels of Fig.~\ref{fig:ppppbpbpbg2cicd}, it is readily noticed   that the $G_2^{\rm CI}$ correlator   features different evolutions with $[N_{\rm ch}]$ in the azimuthal and the longitudinal directions.   
A strong   narrowing with increasing $[N_{\rm ch}]$ is observed for the azimuthal width, $\sigma_{\Delta\varphi}$, in pp and  p--Pb collisions, as well as in Pb--Pb collisions~\cite{ALICE:2019smr}. The magnitude and  the evolution with $[N_{\rm ch}]$ of the azimuthal widths measured in pp and p--Pb collisions are consistent between each other. In contrast, the  width measured in the lowest Pb--Pb multiplicity class (most peripheral collisions), exceeds the widths observed at similar multiplicity in p--Pb collisions by $\approx8\%$ thereby indicating a difference between the correlations established in p--Pb and Pb--Pb collisions with similar charged particle multiplicity. 

The  $[N_{\rm ch}]$ evolution of the longitudinal width of the 
$G_2^{\rm CI}$ correlator is different for the three collision systems and contrasts markedly from the trend measured for the azimuthal width $\sigma_{\Delta\varphi}$. In pp collisions, the width exhibits a trend consistent with a very modest narrowing with increasing $[N_{\rm ch}]$ whereas, in p--Pb collisions, the data suggests a weak increase with  $[N_{\rm ch}]$. It is also found that at equal values of  $[N_{\rm ch}]$ the longitudinal widths measured in p--Pb are also  somewhat larger than those observed in pp collisions even though they are compatible within uncertainties. The increasing trend seen in p--Pb is difficult to precisely assess given the size of the systematic uncertainties relative to the very modest increase of the width. It is rather clear, nonetheless, that it does not match the rapid and large  increase observed in  Pb--Pb collisions. Indeed, at   $[N_{\rm ch}] \approx 50$, the longitudinal width observed in Pb--Pb exceeds that measured in p--Pb collisions by $\approx13\%$. The slope of the increasing trend of $\sigma_{\Delta\eta}$ in Pb--Pb collisions far exceeds that seen in p--Pb. By extrapolating the trend  observed in Pb--Pb  to small $[N_{\rm ch}]$, the obtained $\sigma_{\Delta\eta}$ values match those measured in pp collisions but it is rather clear that the broadening observed in Pb--Pb collisions stands in stark contrast to the evolution observed in the smaller systems.
Overall,   the current measurements indicate that, while the azimuthal widths of the CD and CI correlators in pp and p--Pb collisions show a trend with multiplicity compatible with that found in the larger Pb--Pb system, the evolution of the longitudinal width of the $G_2^{\rm CI}$   correlator is rather different in small and large systems. 

It is of interest to contrast the results of the bidimensional fit procedure used in this work with those obtained with the zero yield at minimum (ZYAM) method~\cite{Trainor:2009gj} widely used in the analysis of azimuthal correlation functions. The difficulty with ZYAM is that if the peaks are wide in $\Delta\varphi$, the gap between the near-side and away-side peaks gets ``filled up'' in the projection. In particular, if the ZYAM method is indiscriminately applied to correlation functions with a strong dependence on the longitudinal particle pair separation, $\Delta\eta$, significant biases may occur in the evaluation of the amplitudes and widths of such correlations and their dependence on the global event observables.

\begin{figure}[ht]
  \centering
  \includegraphics[width=0.49\textwidth,keepaspectratio=true,clip=true,trim=2pt 4pt 43pt 2pt]
  {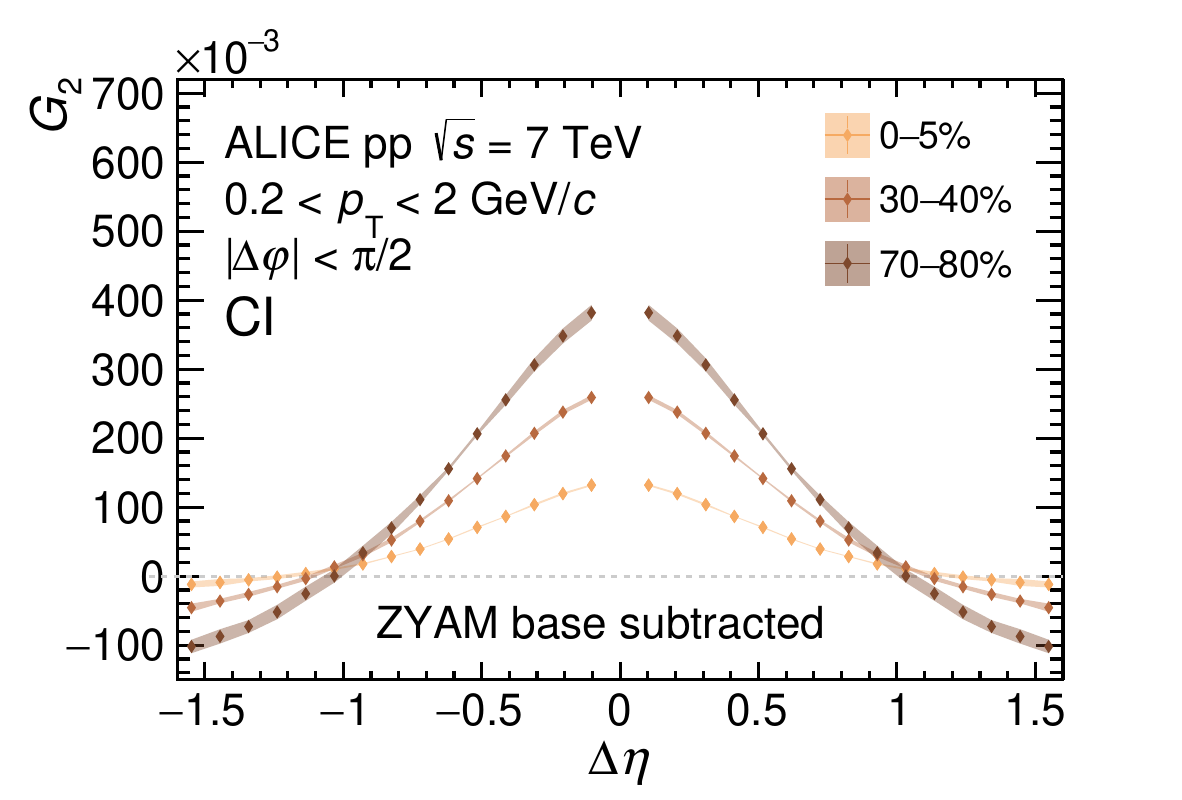}
  \includegraphics[width=0.49\textwidth,keepaspectratio=true,clip=true,trim=2pt 4pt 43pt 2pt]
  {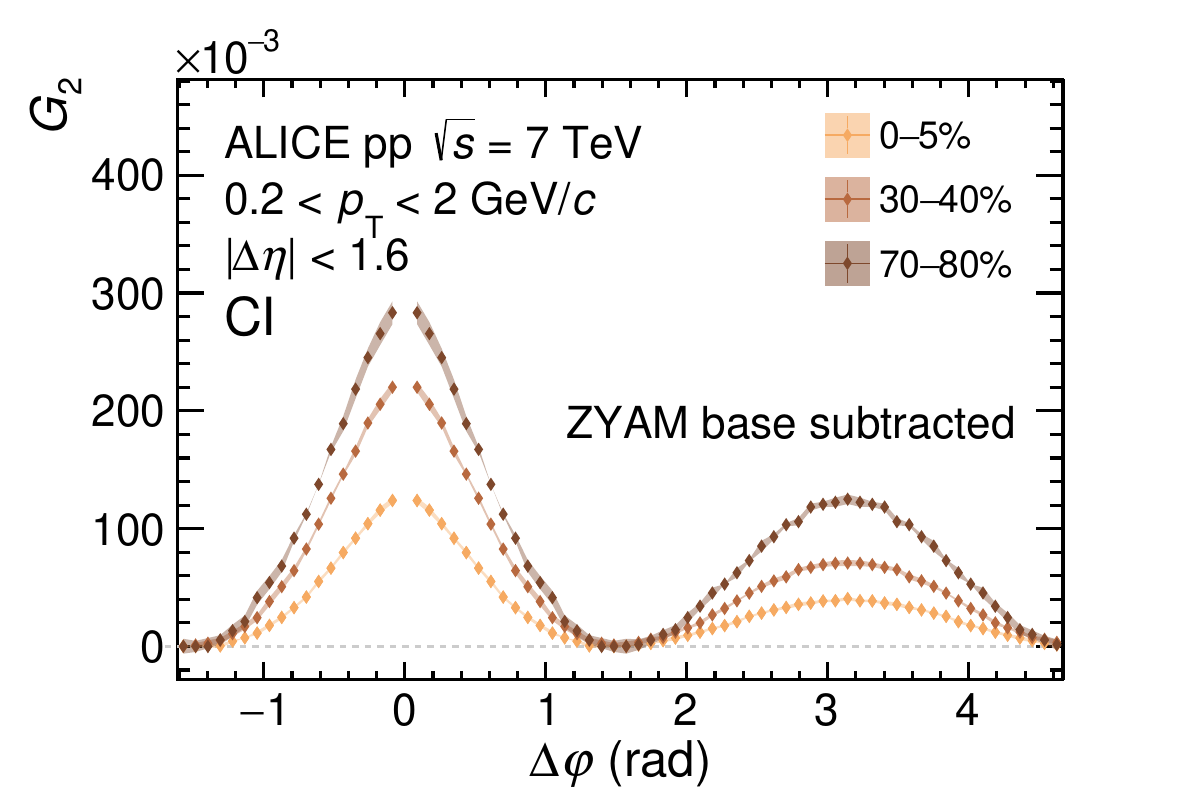} \par
  \caption{\label{fig:zyam} Longitudinal (left) and azimuthal (right) projections of the two-particle transverse momentum correlation $G_{2}^{\rm CI}$ for selected charged particle multiplicity classes in pp collisions at $\sqrt{s}=7\;\text{TeV}$ after subtracting the azimuthal ZYAM base level (see text for details) for each multiplicity class. Vertical bars (mostly smaller than the marker size) and shaded bands represent statistical and uncorrelated systematic uncertainties, respectively.}
\end{figure}

As an illustration, the right panel of  Fig.~\ref{fig:zyam} shows the azimuthal projections of the $G_{2}^{\rm CI}$ correlator for three selected multiplicity classes, after applying the ZYAM procedure to remove ``uncorrelated backgrounds'', i.e.~by uniformly subtracting  a constant value corresponding to the minimum yield value, hereafter called ZYAM base level. The widths extracted based on  these projections are  approximately the same for the different multiplicity classes considered, thereby leading to the  conclusion that the width of these correlation functions is independent of the multiplicity class. In the left panel of Fig.~\ref{fig:zyam}, the near-side longitudinal projections of the $G_{2}^{\rm CI}$ correlator are shown after subtraction of the ZYAM base level. Parts of the longitudinal projections lie significantly below zero, with values that depend on the multiplicity class. The notion of zero yield at minimum is thus rather poorly defined in this context given that the minimum of the longitudinal correlations (within the measurement acceptance) significantly deviates from  the ZYAM base level and is a monotonic function of the multiplicity class. The application of the ZYAM method consequently  does not enable a simultaneous consistent extraction of the azimuthal and longitudinal widths of the $G_2^{\rm CI}$ correlators measured in this work.

\begin{figure}[ht]
  \centering
  \includegraphics[width=0.49\textwidth,keepaspectratio=true,clip=true,trim=2pt 57pt 43pt 2pt]
  {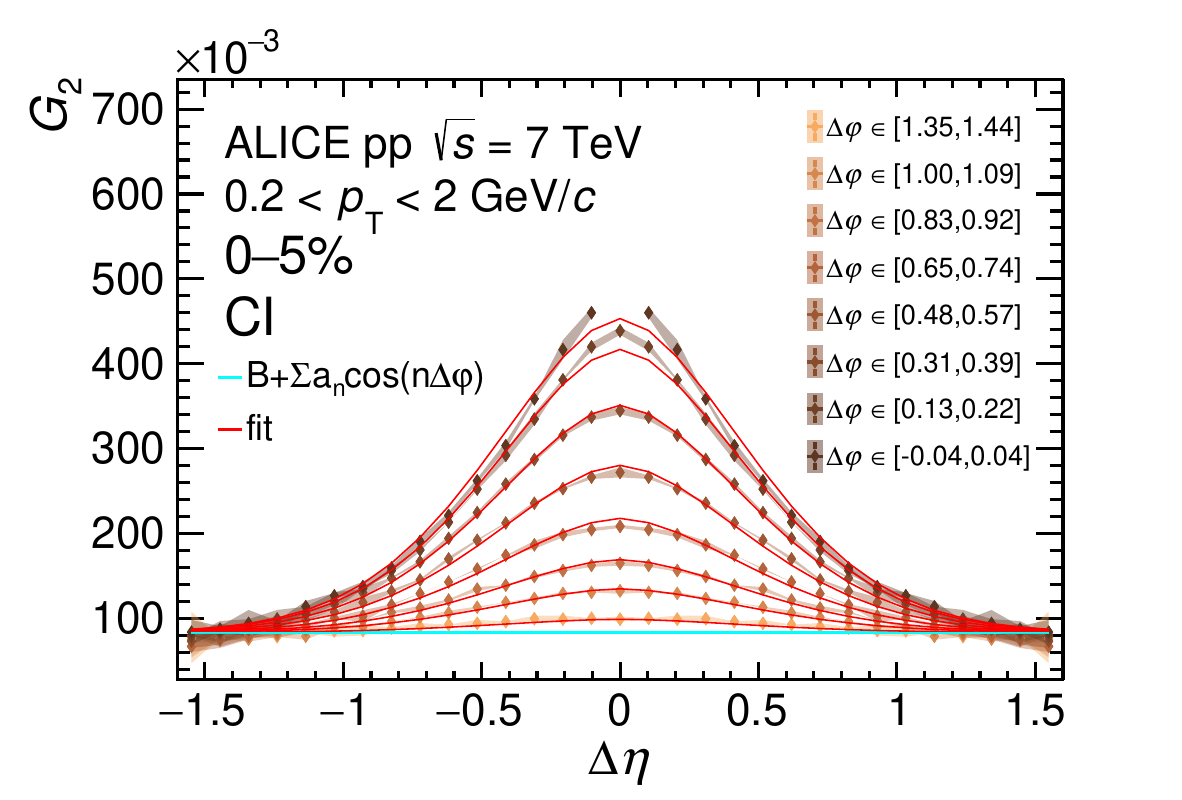}
  \includegraphics[width=0.49\textwidth,keepaspectratio=true,clip=true,trim=2pt 57pt 43pt 2pt]
  {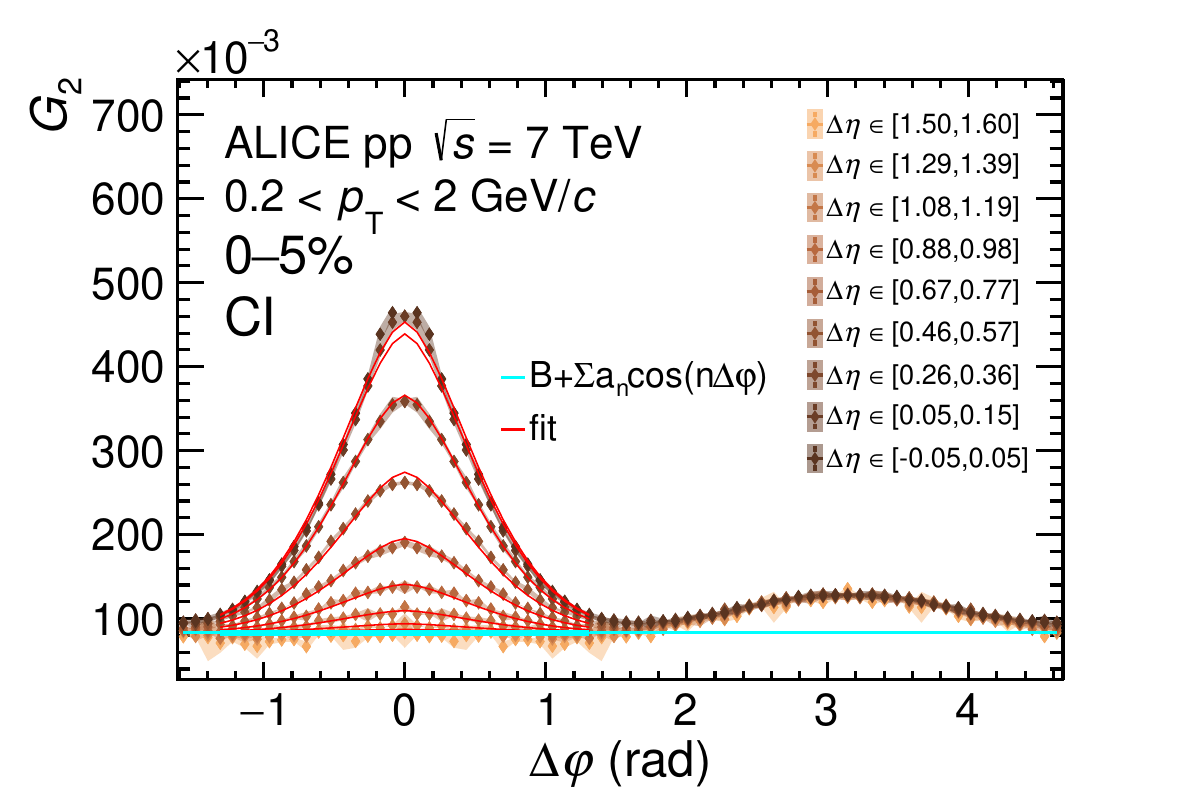}\par
  \includegraphics[width=0.49\textwidth,keepaspectratio=true,clip=true,trim=2pt 4pt 43pt 30pt]
  {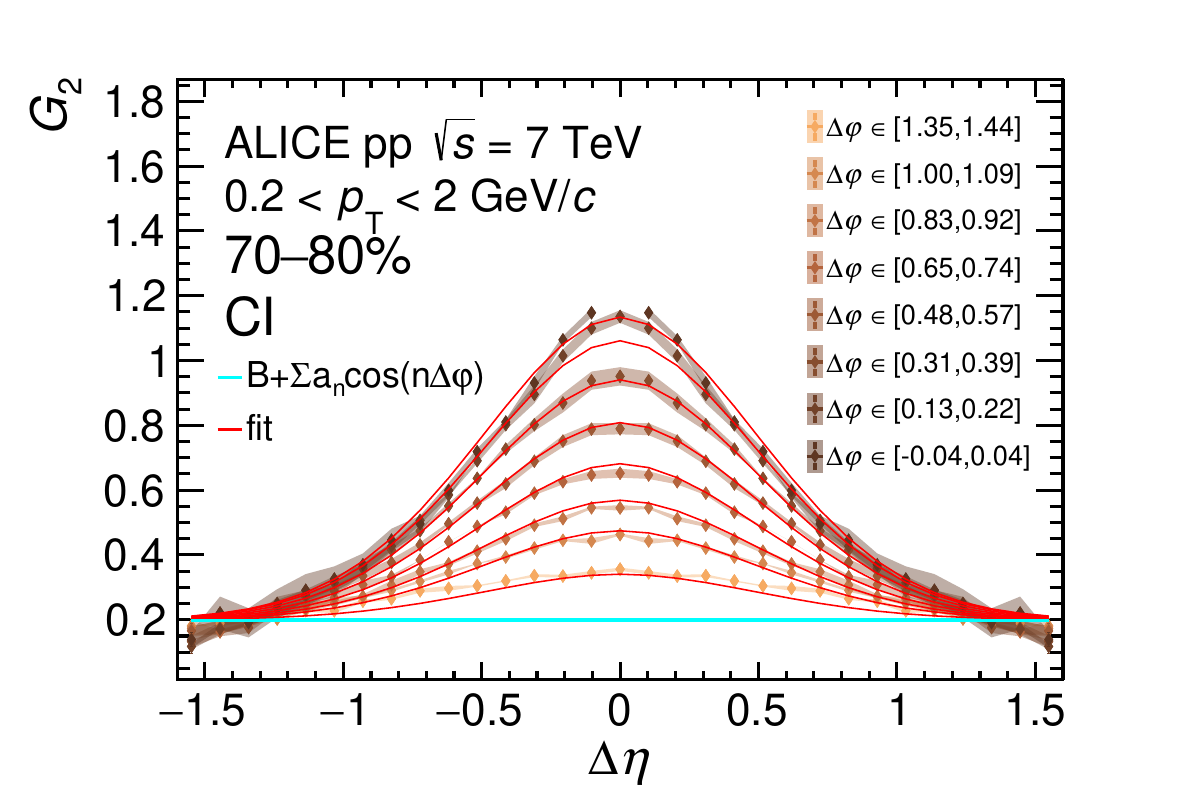}
  \includegraphics[width=0.49\textwidth,keepaspectratio=true,clip=true,trim=2pt 4pt 43pt 30pt]
  {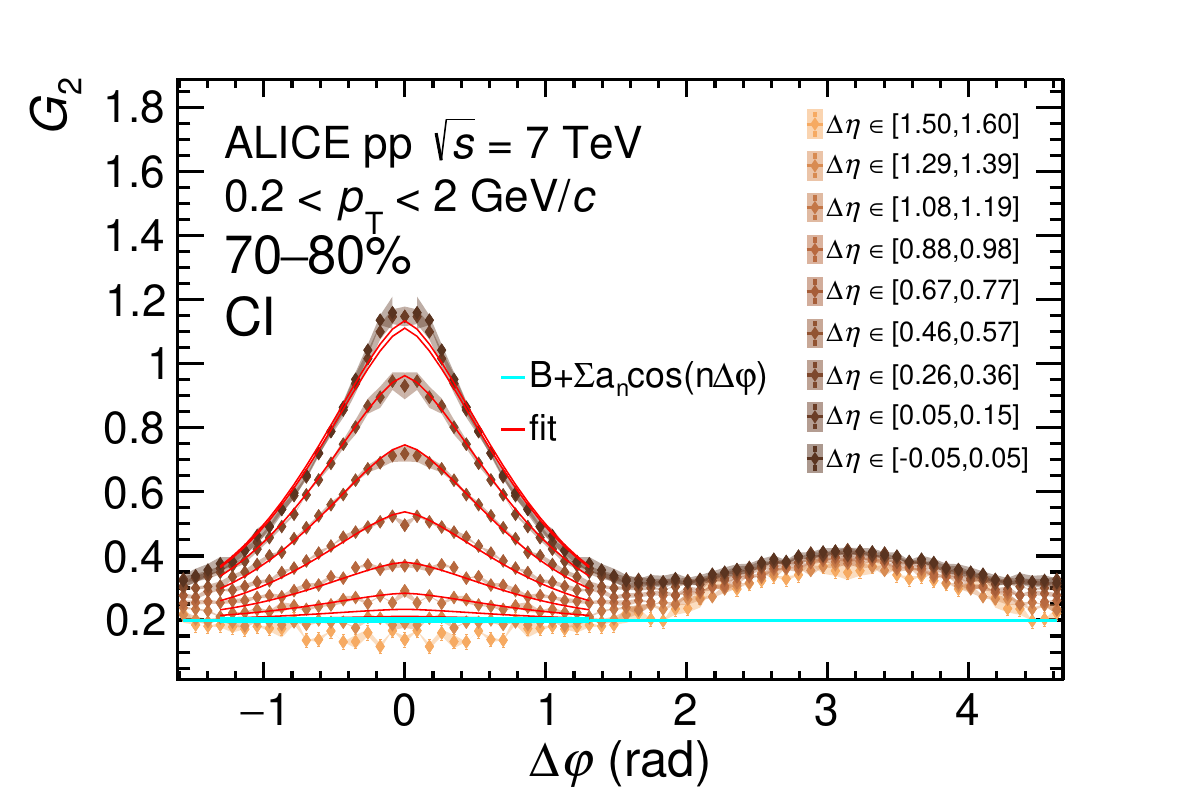}\par
  \caption{\label{fig:fitprojections} Longitudinal projections of slices of one azimuthal bin (left) and azimuthal projections of slices of one longitudinal bin (right), for selected bins of the two-particle transverse momentum correlation $G_{2}^{\rm CI}$ and its bidimensional fit using Eq.~(\ref{eq:fitfunc}) for the 0--5\% (top) and 70--80\% (bottom) charged particle multiplicity classes in pp collisions at $\sqrt{s}=7\;\text{TeV}$. Vertical bars (mostly smaller than the marker size) and shaded bands represent statistical and uncorrelated systematic uncertainties, respectively.}
\end{figure}
By contrast, the fit method used in this work  parameterises the correlation functions with a bidimensional generalised Gaussian  model, Eq.~\ref{eq:fitfunc}, with independent parameters along the $\Delta\eta$ and $\Delta\varphi$ directions, thereby enabling a more accurate description of the shape of the $G_2^{\rm CI}$ correlator and its dependence on the multiplicity $[N_{\rm ch}]$. For illustrative purposes, Fig.~\ref{fig:fitprojections} compares several projections of the two-dimensional fit functions (red lines) and the data (solid diamonds) for the CI correlators in the 0--5\% and 70--80\% multiplicity classes of pp collisions. Projections onto $\Delta\eta$  (left panels) are shown for selected ranges of $\Delta\varphi$ pair separation, and conversely, projections onto $\Delta\varphi$ (right panels) are displayed for selected ranges of $\Delta\eta$ pair separation. 
The cyan lines represent the baseline $B$ plus the anisotropic modulations given by the coefficients $a_{\rm n}$ in the fit function, extended to the whole azimuthal range in the case of the azimuthal projections, in which the thicker line section represents the azimuthal portion considered for the bidimensional fit. The two-dimensional fits provide good descriptions of most of the  azimuthal slices, shown in the left panels, (i.e.~irrespective of the multiplicity class and $\Delta\varphi$ range) as well as good match on most of the longitudinal slices, shown in the right panels. Deviations of the model from data are observed at large longitudinal relative separation, visible for slices at large $\Delta\eta$, and in the proximity of $\Delta\eta, \Delta\varphi = (0,0)$. The two-dimensional generalised Gaussian model  used in this work thus provides a reliable, robust, and self-consistent description of the $G_2^{\rm CI}$ correlator measurements. The azimuthal and longitudinal widths extracted from this model thus do not suffer from the biased and inconsistent behaviour obtained with the ZYAM method. 
The right panels of Fig.~\ref{fig:fitprojections}, displaying the $\Delta\varphi$ projections of $G_2^{\rm CI}$, provide a simple explanation of the bias encountered with the basic ZYAM method. The shape and strength of the away side, $\pi/2 \le \Delta\varphi < 3\pi/2$, are essentially independent of the $\Delta\eta$ range considered whereas the near side, $|\Delta\varphi|<\pi/2$, is strongly  dependent on $\Delta\eta$. 
The ZYAM values of these $\Delta\varphi$ projections  therefore depend on $\Delta\eta$ and the multiplicity classes. This thus results in inconsistent extractions of the longitudinal and azimuthal widths of the correlator if the basic ZYAM method is used. Such issues are clearly avoided
with the two-dimensional generalised Gaussian fit method utilised in this work.

\section{Comparison with results from event generators}
\label{sec:models}

A number of event generators have had great successes in quantitatively reproducing the many features and properties of particle production in pp, p--Pb, and Pb--Pb collisions~\cite{Mitrovski:2008hb,Petersen:2011sb,Basu:2016dmo,Solanki:2012ne,Werner:2007bf,Basu:2020jbk,Knospe:2015nva}. It is thus legitimate to consider whether such  production models can also match the magnitude and the evolution with $[N_{\rm ch}]$ of the near-side peak widths reported in Fig.~\ref{fig:ppppbpbpbg2cicd}. 
Comparisons of calculations of the two-particle number correlator $R_2(\Delta\eta,\Delta\varphi)$ and the two-particle transverse momentum correlator $P_2(\Delta\eta,\Delta\varphi)$ performed  with the AMPT, EPOS, and UrQMD models~\cite{Basu:2020ldt} with data  reported by the ALICE Collaboration~\cite{ALICE:2018jco} show these three event generators are considerably challenged by the measurements. In particular, since these models do not fully implement charge and baryon number conservation, they cannot reproduce the salient features of the measured $R_2^{\rm CD}$ and $P_2^{\rm CD}$ correlators, while they qualitatively reproduce some but not all facets of the measured $R_2^{\rm CI}$ and $P_2^{\rm CI}$ correlation functions.

\begin{figure}[hb]
  \includegraphics[width=0.99\textwidth,keepaspectratio=true,clip=true,trim=0pt 0pt 0pt 0pt]
  {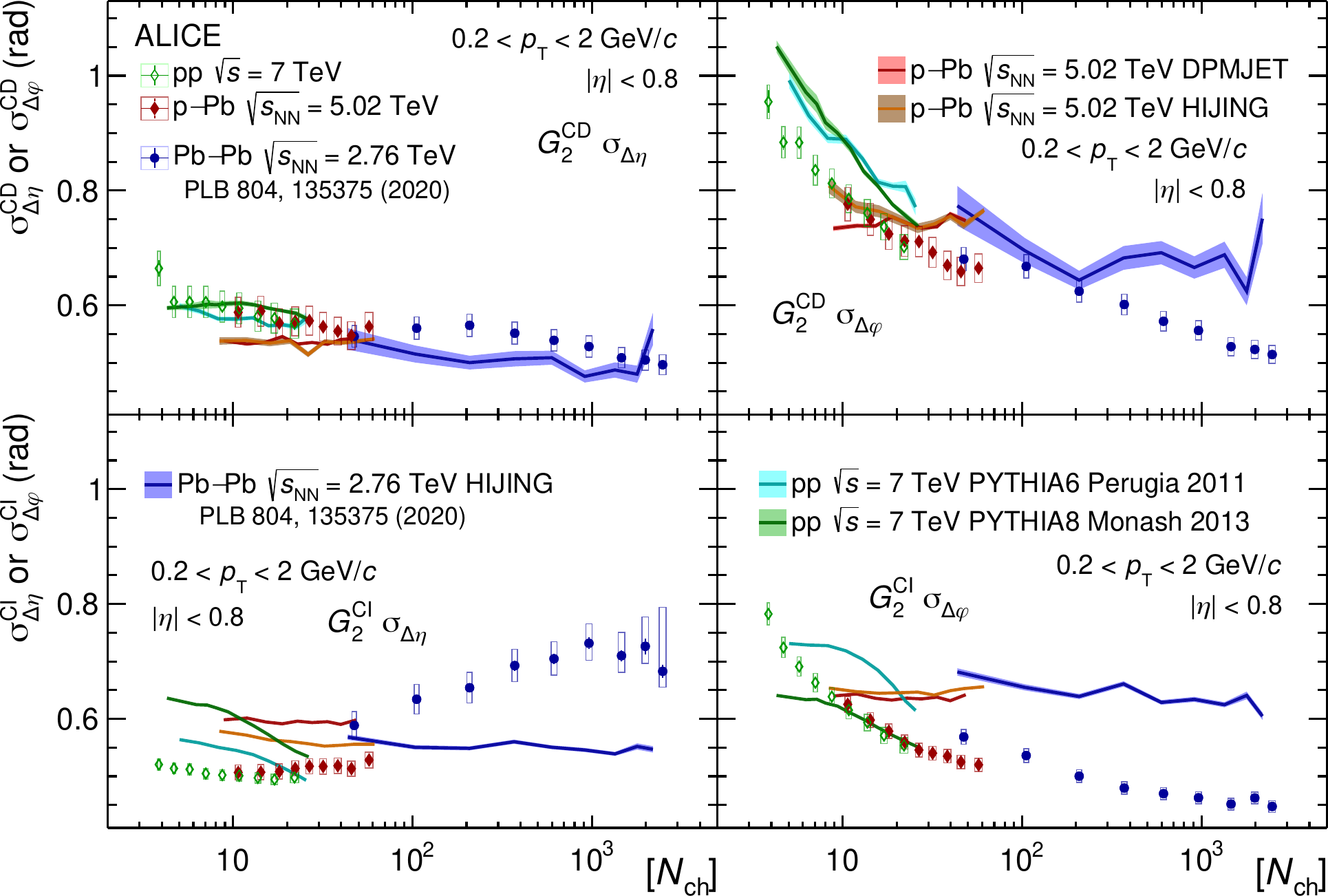}
  \caption{Evolution with the average charged particle multiplicity of the longitudinal (left) and azimuthal (right) widths of the two-particle transverse momentum correlations $G_{2}^{\rm CD}$ (top row) and $G_{2}^{\rm CI}$ (bottom row) in pp, p--Pb, and Pb--Pb collisions at $\sqrt{s} = 7\;\text{TeV}$, $\sqrt{s_{\rm NN}} = 5.02\;\text{TeV}$, and $\sqrt{s_{\rm NN}} = 2.76\;\text{TeV}$, respectively, compared to models. Statistical and systematic uncertainties of the data points are shown as vertical bars (mostly smaller than the marker size) and filled boxes, respectively, while the thickness of the shaded bands represents statistical uncertainties of the models. The data points and the results of HIJING simulations for Pb--Pb collisions are taken from Ref.~\cite{ALICE:2019smr}.}
  \label{fig:ppppbpbpbg2cicdwmodel}
\end{figure}

The discussion in this section is limited to four well established models: PYTHIA~6~\cite{Sjostrand:2006za} (Perugia default tune~\cite{Skands:2010ak}) and PYTHIA~8~\cite{Sjostrand:2007gs} (Monash tune, with colour reconnection~\cite{Skands:2014pea}) for comparison with pp data, DPMJET~\cite{Roesler:2000he} for comparison with p--Pb data, and HIJING~\cite{Wang:1991hta} for comparison with both p--Pb and Pb--Pb data. PYTHIA and DPMJET are known to well reproduce measurements of  differential cross section in pp collisions and, although HIJING does not   include a modelling of the collective behaviour observed in Pb--Pb, it is here used as a baseline reference for the discussion of trends as a function of multiplicity in that system.
Simulated data sets produced with these four event generators are  analysed, at generator level,  with identical event and charged particle selection criteria and multiplicity classes as the data. This allows to obtain the $G_2^{\rm CD}(\Delta\eta,\Delta\varphi)$ and $G_2^{\rm CI}(\Delta\eta,\Delta\varphi)$ correlation functions, which are then fitted with Eq.~(\ref{eq:fitfunc}). The width parameters obtained from the fits to the  simulated  correlators 
are compared to the measured ones in Fig.~\ref{fig:ppppbpbpbg2cicdwmodel}.
The data from Pb--Pb collisions and the results of simulations with HIJING are taken from Ref.~\cite{ALICE:2019smr}.

In the case of the charge-dependent correlator  $G_{2}^{\rm CD}$ shown in the top panels of Fig.~\ref{fig:ppppbpbpbg2cicdwmodel}, the measured magnitude and multiplicity dependence of the longitudinal width of the near-side peak, $\sigma_{\Delta\eta}^{\rm CD}$, are described within uncertainties by both PYTHIA~6 and PYTHIA~8 simulations. The two PYTHIA tunes  also qualitatively reproduce the observed narrowing trend of the  azimuthal  width of $G_{2}^{\rm CD}$ as a function of $[N_{\rm ch}]$ even though they overestimate the magnitude of $\sigma_{\Delta\varphi}$.
By contrast, although   DPMJET and  HIJING qualitatively reproduce the magnitudes of the longitudinal and azimuthal widths of the $G_{2}^{\rm CD}$ correlator, they have rather limited success in describing  the evolution with multiplicity of the azimuthal widths  observed  in p--Pb  and Pb--Pb collisions. 

In the bottom panels of Fig.~\ref{fig:ppppbpbpbg2cicdwmodel}, the measured evolution of the $G_{2}^{\rm CI}$ correlator widths, $\sigma_{\Delta\eta}^{\rm CI}$ and $\sigma_{\Delta\varphi}^{\rm CI}$, with multiplicity is compared to model predictions. PYTHIA~8 describes well the measured azimuthal widths $\sigma_{\Delta\varphi}$ for large multiplicities, but it underestimates them for low multiplicity pp collisions~($[N_{\rm ch}] < 10$). Instead PYTHIA~6 grossly overestimates the width and misses the observed trend as a function of $[N_{\rm ch}]$. Neither of the two PYTHIA versions successfully reproduces the $[N_{\rm ch}]$ evolution of $\sigma_{\Delta\eta}^{\rm CI}$, although PYTHIA~6 quantitatively matches the width observed in the highest pp multiplicity class. Similarly, predictions by both DPMJET and HIJING clearly overestimate the longitudinal and azimuthal widths measured in p--Pb collisions. Although HIJING manages to approximately match the longitudinal width observed at lowest multiplicity in Pb--Pb collisions, it  systematically underpredicts the $\sigma_{\Delta\eta}$ values measured in higher Pb--Pb multiplicity classes and thus fails to match the broadening trend of $\sigma_{\Delta\eta}^{\rm CI}$ vs. $[N_{\rm ch}]$. Overall, the models considered for the comparison match the evolution of the longitudinal and azimuthal widths of $G_2^{\rm CD}$ the closest while doing rather poorly for the evolution of the widths of $G_2^{\rm CI}$. This suggests that their ability to describe charge conserving processes is decent but that, globally, some new features are needed to properly match the evolution of the widths of $G_2^{\rm CI}$.

\section{Discussion}
\label{sec:discussion}

The inspection of the evolution of the $G_{2}^{\rm CD}$ and $G_{2}^{\rm CI}$ correlators with the multiplicity of charged particles produced in the collision suggests that two or more competing mechanisms may be at play in pp, p--Pb, and Pb--Pb collisions: while the near side peak of $G_{2}^{\rm CD}$ shows narrowing trends with increasing $[N_{\rm ch}]$ in these systems, $G_{2}^{\rm CI}$ exhibits mixed trends in the longitudinal dimension, a strong broadening in Pb--Pb, minor broadening in p--Pb, and modest narrowing in pp collisions. 

A narrowing of the near-side peak of two-particle correlators has  been observed for $R_2^{\rm CD}$ and  $P_2^{\rm CD}$ differential correlators of low-$p_{\rm T}$ particles measured  in Pb--Pb collisions at $\sqrt{s_{\rm NN}} = 2.76$ TeV~\cite{ALICE:2018jco}, for balance functions of charged particles and identified hadrons in pp, p--Pb, and Pb--Pb collisions at 7, 5.02, and 2.76 TeV, respectively~\cite{ALICE:2013vrb, ALICE:2015nuz, ALICE:2021hjb}, as well as in Au--Au collisions at RHIC energies~\cite{STAR:2003kbb,STAR:2010plm,STAR:2015ryu}. 
Measurements in large collision systems show that the average transverse momentum of produced particles, $\langle p_{\rm T} \rangle$, rises monotonically with increasing multiplicity in   Au--Au and Pb--Pb collisions and saturates in most central collisions. Similarly, blast-wave model fits of $p_{\rm T}$ spectra measured  as a function of increasing collision centrality in these systems  indicate that the radial flow velocity increases with increasing multiplicity of produced particles~\cite{Retiere:2003kf,ALICE:2013mez}. Moreover, theoretical studies of the evolution of the narrowing of charge balance functions, $B_{\rm ch}$, in Au--Au collisions based on the blast-wave model   are found to require  strong radial flow to match the progressive narrowing of the near-side peak of these correlation functions~\cite{Schlichting:2010qia}. Similar narrowing effects and large radial flow are also seen in   hydrodynamics calculations~\cite{Pratt:2021xvg}.

Given that $G_2$ shares the same correlation kernel as the $R_2$, $P_2$, and $B_{\rm ch}$ correlators, the progressive narrowing of its near-side peak with increasing values of $[N_{\rm ch}]$ in Pb--Pb collisions can be interpreted as  resulting from the increasing radial flow velocity. An increase of $\langle p_{\rm T} \rangle$ with increasing multiplicity is also observed  in pp and p--Pb collisions~\cite{Abelev:2013bla}. However, it is still an open question whether this increase results from the production of a radially flowing medium also in these small collision systems. In any case, it is expected that an  increase in $\langle p_{\rm T} \rangle$ should also produce  a kinematic focusing of correlated particles, resulting in a narrowing of the correlation peaks. A modest  narrowing of the $R_2^{\rm CD}$ and $P_2^{\rm CD}$ correlation functions  has already been reported in p--Pb collisions~\cite{ALICE:2018jco} and is, with this work, also established for the azimuthal widths of  $G_{2}^{\rm CD}$ and $G_{2}^{\rm CI}$ correlators observed in the pp and p--Pb systems. It is  then plausible to postulate that the azimuthal narrowing of the correlators results from collective radial flow in these smaller systems.

The details of the evolution of the widths $\sigma_{\Delta\varphi}^{\rm CD}$ and $\sigma_{\Delta\varphi}^{\rm CI}$ with  $[N_{\rm ch}]$  are consequently  of particular interest. As already suggested, both these widths exhibit a monotonic narrowing with increasing $[N_{\rm ch}]$ in all three systems studied. However, their evolution is not continuous between systems. Indeed, at equal particle production  $[N_{\rm ch}]$,  the width of these correlators in p--Pb collisions is similar, within uncertainties, to that observed in pp collisions. 
Additionally, although measurements in p--Pb and Pb--Pb collisions have a  limited overlap in $[N_{\rm ch}]$, the $\sigma_{\Delta\varphi}^{\rm CI}$  widths measured in Pb--Pb are approximately 8\% larger than those observed in p--Pb. These differences can  qualitatively be  interpreted as resulting from the size of the systems considered. The lowest multiplicity measurement of $G_2$ in Pb--Pb collisions shown in Fig.~\ref{fig:ppppbpbpbg2cicd} corresponds to the  70--80\% collision centrality class. In this range,  the average  number of nucleons participating in the Pb--Pb collision amounts to approximately 15~\cite{Loizides:2017ack} and should far exceed the number of participants involved in a typical  p--Pb collision. Although estimates of the number of participant nucleons in p--Pb collisions are less precise than in the Pb--Pb case, it can be reasonably expected that at a given value of $[N_{\rm ch}]$ p--Pb collisions involve on average more participants than  pp collisions. At a given value of $[N_{\rm ch}]$,  p--Pb collisions (Pb--Pb collisions) are thus expected to consist of a lower number of contributors from hard scatterings than those produced in  pp (p--Pb) collisions. It is then reasonable to expect that  these correlators in different collision systems have slightly different widths at equal $[N_{\rm ch}]$ value. In any case, comprehensive models of particle production in small and large systems should account for these small differences, and the evolution of the azimuthal width of $G_2^{\rm CD}$ thus constitutes a stringent test of such models.

The three systems exhibit a rather different evolution of the longitudinal width   $\sigma_{\Delta\eta}^{\rm CI}$ with increasing $[N_{\rm ch}]$. Whereas $\sigma_{\Delta\eta}^{\rm CI}$ increases from 0.59 to a maximum value of 0.73 (24\%) with increasing $[N_{\rm ch}]$ in Pb--Pb collisions,  it  decreases by about 5\% in pp while rising by approximately the same amount in p--Pb collisions. A broadening of the longitudinal width of the near-side peak in the $G_2^{\rm CI}$ correlator was also observed in Au--Au collisions at RHIC~\cite{STAR:2011iio} and was predicted to occur as a result of viscous forces in collision systems producing a long lived QGP phase~\cite{Gavin:2006xd}. In this context, the medium formed in the collision is modeled as a fluid in quasi-equilibrium. Fluid cells are accelerated by local pressure gradients and do not, ab initio, have equal transverse  velocities. Viscous interactions between cells are then expected to slow down fast moving cells and accelerate cells with smaller transverse velocity. The resulting momentum transfers are  then expected to  induce  ``long range'' longitudinal correlations between cells and the particles they emit,  thereby producing broadened two-particle correlation functions relative to systems that do not undergo viscous forces. 

The Pb--Pb system clearly manifests the behaviour expected from a long lived viscous fluid: the average transverse momentum monotonically increases with increasing multiplicity of produced particles, the width $\sigma_{\Delta\varphi}^{\rm CD}$ monotonically decreases, whereas the width $\sigma_{\Delta\eta}^{\rm CI}$ increases with increasing multiplicity and is consistent with a fluid characterised by a small value of  $\eta/s$~\cite{ALICE:2019smr,Gonzalez:2020bqm}. 

The behaviour and nature  of the pp and p--Pb systems are much less clear. While both collisions systems exhibit the $\sigma_{\Delta\varphi}^{\rm CD}$ narrowing expected from a fluid undergoing radial flow and  anisotropic particle emission in the transverse plane, the width $\sigma_{\Delta\eta}^{\rm CI}$ decreases with increasing multiplicity in pp collisions while exhibiting a very modest increase in p--Pb collisions.  
No conclusion can therefore be drawn on the possible establishment of a collective behaviour in pp collisions, while the results from p--Pb collisions provide only a suggestive indication, limited by the magnitude of systematic uncertainties, for the presence of viscous effects. The interpretation of trends observed in these two collision systems is further complicated by the presence of competing effects resulting from radial flow. 
The longitudinal widths $\sigma_{\Delta\eta}^{\rm CD}$ in pp and p--Pb collisions exhibit a narrowing trend with increasing $[N_{\rm ch}]$ that should  be matched, in the absence of viscous effects,  by a similar behaviour for $\sigma_{\Delta\eta}^{\rm CI}$. Such small  narrowing effect may then compete with and partly mask the viscous broadening  that would otherwise occur in such  small systems. It should also be considered that possible viscous forces  would need time to propagate correlations in the longitudinal direction. Thus, the longitudinal broadening  of $G_{2}$ shall depend both on the magnitude of the shear viscosity (per unit entropy) and  the lifetime of the fluid. If the fluid-like system produced is too small or too short lived, there may not be enough time for viscous forces to equalise the transverse velocity differences between cells and, even though a fluid-like system may be produced in pp or p--Pb collisions, it may not live long enough to yield a significant broadening of the $G_{2}^{\rm CI}$ correlator. Competing effects associated with kinematic focusing may then hinder observations of viscous broadening. 

Alternatively, it is also possible that a quasi-equilibrated fluid description does not hold for the system produced in pp and p--Pb collisions. Appealing to more traditional models to interpret the data is then needed. However, as  already noted, while PYTHIA~6 manages to qualitatively reproduce the narrowing of the CD correlator, it poorly describes the measurements for the CI correlator. Similarly, PYTHIA~8 (Monash tune, with  colour reconnection) qualitatively reproduces  the azimuthal widths but introduces too much narrowing in the longitudinal dimension of the CI correlator. 

\section{Conclusions}
\label{sec:conclusions}
The two-particle transverse momentum differential correlators $G_{2}^{\rm CI}$ and $G_{2}^{\rm CD}$ were measured in pp collisions at $\sqrt{s} = 7\;\text{TeV}$ and in p--Pb collisions at $\sqrt{s_{\rm NN}} = 5.02\;\text{TeV}$ as a function of the charged hadron multiplicity measured in the selected acceptance, $[N_{\rm ch}]$. Both correlators feature prominent near-side peaks. 
The amplitude of these  peaks    decreases monotonically with increasing  charged hadron multiplicity, $N_{\rm ch}$, in both collision systems, but their widths exhibit mixed behaviours. 

The near-side peak of the $G_{2}^{\rm CI}$ and $G_{2}^{\rm CD}$  correlators exhibits  strong azimuthal narrowing trends with increasing $N_{\rm ch}$, in pp and p--Pb collisions, that qualitatively match the width evolution with collision centrality formerly observed in the Pb--Pb system. The $G_{2}^{\rm CD}$ near-side peak also features a longitudinal narrowing albeit weaker than that observed in the azimuthal direction. The narrowing  trends observed in pp collisions are qualitatively reproduced by PYTHIA, even  with PYTHIA 6 Perugia tune, thereby indicating  that the evolution of the two correlators is well accounted for by this model, i.e. without the need to invoke a collective behaviour. However, the multiplicity dependence measured in p--Pb collisions is not described by DPMJET.

The longitudinal width, $\sigma_{\Delta\eta}^{\rm CI}$, of the $G_{2}^{\rm CI}$ correlator in pp and p--Pb collisions is not varying with multiplicity, within uncertainties, at variance with the case of Pb--Pb collisions. The lack of a clear dependence on multiplicity of the widths in pp and p--Pb collisions provides no evidence of an increase of $\sigma_{\Delta\eta}^{\rm CI}$ within uncertainties. 
It is possible that, if fluid-like systems are produced in p--Pb collisions, they are too short lived for viscous forces to have a sizable impact on the width of the correlator. Further studies of the p--Pb system are thus required to fully elucidate its behaviour.


\newenvironment{acknowledgement}{\relax}{\relax}
\begin{acknowledgement}
\section*{Acknowledgements}

The ALICE Collaboration would like to thank all its engineers and technicians for their invaluable contributions to the construction of the experiment and the CERN accelerator teams for the outstanding performance of the LHC complex.
The ALICE Collaboration gratefully acknowledges the resources and support provided by all Grid centres and the Worldwide LHC Computing Grid (WLCG) collaboration.
The ALICE Collaboration acknowledges the following funding agencies for their support in building and running the ALICE detector:
A. I. Alikhanyan National Science Laboratory (Yerevan Physics Institute) Foundation (ANSL), State Committee of Science and World Federation of Scientists (WFS), Armenia;
Austrian Academy of Sciences, Austrian Science Fund (FWF): [M 2467-N36] and Nationalstiftung f\"{u}r Forschung, Technologie und Entwicklung, Austria;
Ministry of Communications and High Technologies, National Nuclear Research Center, Azerbaijan;
Conselho Nacional de Desenvolvimento Cient\'{\i}fico e Tecnol\'{o}gico (CNPq), Financiadora de Estudos e Projetos (Finep), Funda\c{c}\~{a}o de Amparo \`{a} Pesquisa do Estado de S\~{a}o Paulo (FAPESP) and Universidade Federal do Rio Grande do Sul (UFRGS), Brazil;
Bulgarian Ministry of Education and Science, within the National Roadmap for Research Infrastructures 2020¿2027 (object CERN), Bulgaria;
Ministry of Education of China (MOEC) , Ministry of Science \& Technology of China (MSTC) and National Natural Science Foundation of China (NSFC), China;
Ministry of Science and Education and Croatian Science Foundation, Croatia;
Centro de Aplicaciones Tecnol\'{o}gicas y Desarrollo Nuclear (CEADEN), Cubaenerg\'{\i}a, Cuba;
Ministry of Education, Youth and Sports of the Czech Republic, Czech Republic;
The Danish Council for Independent Research | Natural Sciences, the VILLUM FONDEN and Danish National Research Foundation (DNRF), Denmark;
Helsinki Institute of Physics (HIP), Finland;
Commissariat \`{a} l'Energie Atomique (CEA) and Institut National de Physique Nucl\'{e}aire et de Physique des Particules (IN2P3) and Centre National de la Recherche Scientifique (CNRS), France;
Bundesministerium f\"{u}r Bildung und Forschung (BMBF) and GSI Helmholtzzentrum f\"{u}r Schwerionenforschung GmbH, Germany;
General Secretariat for Research and Technology, Ministry of Education, Research and Religions, Greece;
National Research, Development and Innovation Office, Hungary;
Department of Atomic Energy Government of India (DAE), Department of Science and Technology, Government of India (DST), University Grants Commission, Government of India (UGC) and Council of Scientific and Industrial Research (CSIR), India;
National Research and Innovation Agency - BRIN, Indonesia;
Istituto Nazionale di Fisica Nucleare (INFN), Italy;
Japanese Ministry of Education, Culture, Sports, Science and Technology (MEXT) and Japan Society for the Promotion of Science (JSPS) KAKENHI, Japan;
Consejo Nacional de Ciencia (CONACYT) y Tecnolog\'{i}a, through Fondo de Cooperaci\'{o}n Internacional en Ciencia y Tecnolog\'{i}a (FONCICYT) and Direcci\'{o}n General de Asuntos del Personal Academico (DGAPA), Mexico;
Nederlandse Organisatie voor Wetenschappelijk Onderzoek (NWO), Netherlands;
The Research Council of Norway, Norway;
Commission on Science and Technology for Sustainable Development in the South (COMSATS), Pakistan;
Pontificia Universidad Cat\'{o}lica del Per\'{u}, Peru;
Ministry of Education and Science, National Science Centre and WUT ID-UB, Poland;
Korea Institute of Science and Technology Information and National Research Foundation of Korea (NRF), Republic of Korea;
Ministry of Education and Scientific Research, Institute of Atomic Physics, Ministry of Research and Innovation and Institute of Atomic Physics and University Politehnica of Bucharest, Romania;
Ministry of Education, Science, Research and Sport of the Slovak Republic, Slovakia;
National Research Foundation of South Africa, South Africa;
Swedish Research Council (VR) and Knut \& Alice Wallenberg Foundation (KAW), Sweden;
European Organization for Nuclear Research, Switzerland;
Suranaree University of Technology (SUT), National Science and Technology Development Agency (NSTDA), Thailand Science Research and Innovation (TSRI) and National Science, Research and Innovation Fund (NSRF), Thailand;
Turkish Energy, Nuclear and Mineral Research Agency (TENMAK), Turkey;
National Academy of  Sciences of Ukraine, Ukraine;
Science and Technology Facilities Council (STFC), United Kingdom;
National Science Foundation of the United States of America (NSF) and United States Department of Energy, Office of Nuclear Physics (DOE NP), United States of America.
In addition, individual groups or members have received support from:
Marie Sk\l{}odowska Curie, European Research Council, Strong 2020 - Horizon 2020 (grant nos. 950692, 824093, 896850), European Union;
Academy of Finland (Center of Excellence in Quark Matter) (grant nos. 346327, 346328), Finland;
Programa de Apoyos para la Superaci\'{o}n del Personal Acad\'{e}mico, UNAM, Mexico.

\end{acknowledgement}

\bibliographystyle{utphys}   
\bibliography{bibliography}

\newpage
\appendix

%
%

\section{The ALICE Collaboration}
\label{app:collab}
\begin{flushleft} 
\small

S.~Acharya\,\orcidlink{0000-0002-9213-5329}\,$^{\rm 125}$, 
D.~Adamov\'{a}\,\orcidlink{0000-0002-0504-7428}\,$^{\rm 86}$, 
A.~Adler$^{\rm 69}$, 
G.~Aglieri Rinella\,\orcidlink{0000-0002-9611-3696}\,$^{\rm 32}$, 
M.~Agnello\,\orcidlink{0000-0002-0760-5075}\,$^{\rm 29}$, 
N.~Agrawal\,\orcidlink{0000-0003-0348-9836}\,$^{\rm 50}$, 
Z.~Ahammed\,\orcidlink{0000-0001-5241-7412}\,$^{\rm 132}$, 
S.~Ahmad\,\orcidlink{0000-0003-0497-5705}\,$^{\rm 15}$, 
S.U.~Ahn\,\orcidlink{0000-0001-8847-489X}\,$^{\rm 70}$, 
I.~Ahuja\,\orcidlink{0000-0002-4417-1392}\,$^{\rm 37}$
% ...

\section*{Affiliation Notes}

$^{\rm I}$ Deceased\\
$^{\rm II}$ Also at: Max-Planck-Institut f\"{u}r Physik, Munich, Germany\\
$^{\rm III}$ Also at: Italian National Agency for New Technologies, Energy and Sustainable Economic Development (ENEA), Bologna, Italy\\
$^{\rm IV}$ Also at: Dipartimento DET del Politecnico di Torino, Turin, Italy\\
$^{\rm V}$ Also at: Department of Applied Physics, Aligarh Muslim University, Aligarh, India\\
$^{\rm VI}$ Also at: Institute of Theoretical Physics, University of Wroclaw, Poland\\
$^{\rm VII}$ Also at: An institution covered by a cooperation agreement with CERN\\

\section*{Collaboration Institutes}

$^{1}$ A.I. Alikhanyan National Science Laboratory (Yerevan Physics Institute) Foundation, Yerevan, Armenia\\
$^{2}$ AGH University of Science and Technology, Cracow, Poland\\
$^{3}$ Bogolyubov Institute for Theoretical Physics, National Academy of Sciences of Ukraine, Kiev, Ukraine\\
$^{4}$ Bose Institute, Department of Physics  and Centre for Astroparticle Physics and Space Science (CAPSS), Kolkata, India\\
$^{5}$ California Polytechnic State University, San Luis Obispo, California, United States\\
$^{6}$ Central China Normal University, Wuhan, China\\
$^{7}$ Centro de Aplicaciones Tecnol\'{o}gicas y Desarrollo Nuclear (CEADEN), Havana, Cuba\\
$^{8}$ Centro de Investigaci\'{o}n y de Estudios Avanzados (CINVESTAV), Mexico City and M\'{e}rida, Mexico\\
$^{9}$ Chicago State University, Chicago, Illinois, United States\\
$^{10}$ China Institute of Atomic Energy, Beijing, China\\
$^{11}$ Chungbuk National University, Cheongju, Republic of Korea\\
$^{12}$ Comenius University Bratislava, Faculty of Mathematics, Physics and Informatics, Bratislava, Slovak Republic\\
$^{13}$ COMSATS University Islamabad, Islamabad, Pakistan\\
$^{14}$ Creighton University, Omaha, Nebraska, United States\\
$^{15}$ Department of Physics, Aligarh Muslim University, Aligarh, India\\
$^{16}$ Department of Physics, Pusan National University, Pusan, Republic of Korea\\
$^{17}$ Department of Physics, Sejong University, Seoul, Republic of Korea\\
$^{18}$ Department of Physics, University of California, Berkeley, California, United States\\
$^{19}$ Department of Physics, University of Oslo, Oslo, Norway\\
$^{20}$ Department of Physics and Technology, University of Bergen, Bergen, Norway\\
$^{21}$ Dipartimento di Fisica, Universit\`{a} di Pavia, Pavia, Italy\\
$^{22}$ Dipartimento di Fisica dell'Universit\`{a} and Sezione INFN, Cagliari, Italy\\
$^{23}$ Dipartimento di Fisica dell'Universit\`{a} and Sezione INFN, Trieste, Italy\\
$^{24}$ Dipartimento di Fisica dell'Universit\`{a} and Sezione INFN, Turin, Italy\\
$^{25}$ Dipartimento di Fisica e Astronomia dell'Universit\`{a} and Sezione INFN, Bologna, Italy\\
$^{26}$ Dipartimento di Fisica e Astronomia dell'Universit\`{a} and Sezione INFN, Catania, Italy\\
$^{27}$ Dipartimento di Fisica e Astronomia dell'Universit\`{a} and Sezione INFN, Padova, Italy\\
$^{28}$ Dipartimento di Fisica `E.R.~Caianiello' dell'Universit\`{a} and Gruppo Collegato INFN, Salerno, Italy\\
$^{29}$ Dipartimento DISAT del Politecnico and Sezione INFN, Turin, Italy\\
$^{30}$ Dipartimento di Scienze MIFT, Universit\`{a} di Messina, Messina, Italy\\
$^{31}$ Dipartimento Interateneo di Fisica `M.~Merlin' and Sezione INFN, Bari, Italy\\
$^{32}$ European Organization for Nuclear Research (CERN), Geneva, Switzerland\\
$^{33}$ Faculty of Electrical Engineering, Mechanical Engineering and Naval Architecture, University of Split, Split, Croatia\\
$^{34}$ Faculty of Engineering and Science, Western Norway University of Applied Sciences, Bergen, Norway\\
$^{35}$ Faculty of Nuclear Sciences and Physical Engineering, Czech Technical University in Prague, Prague, Czech Republic\\
$^{36}$ Faculty of Physics, Sofia University, Sofia, Bulgaria\\
$^{37}$ Faculty of Science, P.J.~\v{S}af\'{a}rik University, Ko\v{s}ice, Slovak Republic\\
$^{38}$ Frankfurt Institute for Advanced Studies, Johann Wolfgang Goethe-Universit\"{a}t Frankfurt, Frankfurt, Germany\\
$^{39}$ Fudan University, Shanghai, China\\
$^{40}$ Gangneung-Wonju National University, Gangneung, Republic of Korea\\
$^{41}$ Gauhati University, Department of Physics, Guwahati, India\\
$^{42}$ Helmholtz-Institut f\"{u}r Strahlen- und Kernphysik, Rheinische Friedrich-Wilhelms-Universit\"{a}t Bonn, Bonn, Germany\\
$^{43}$ Helsinki Institute of Physics (HIP), Helsinki, Finland\\
$^{44}$ High Energy Physics Group,  Universidad Aut\'{o}noma de Puebla, Puebla, Mexico\\
$^{45}$ Horia Hulubei National Institute of Physics and Nuclear Engineering, Bucharest, Romania\\
$^{46}$ Indian Institute of Technology Bombay (IIT), Mumbai, India\\
$^{47}$ Indian Institute of Technology Indore, Indore, India\\
$^{48}$ INFN, Laboratori Nazionali di Frascati, Frascati, Italy\\
$^{49}$ INFN, Sezione di Bari, Bari, Italy\\
$^{50}$ INFN, Sezione di Bologna, Bologna, Italy\\
$^{51}$ INFN, Sezione di Cagliari, Cagliari, Italy\\
$^{52}$ INFN, Sezione di Catania, Catania, Italy\\
$^{53}$ INFN, Sezione di Padova, Padova, Italy\\
$^{54}$ INFN, Sezione di Pavia, Pavia, Italy\\
$^{55}$ INFN, Sezione di Torino, Turin, Italy\\
$^{56}$ INFN, Sezione di Trieste, Trieste, Italy\\
$^{57}$ Inha University, Incheon, Republic of Korea\\
$^{58}$ Institute for Gravitational and Subatomic Physics (GRASP), Utrecht University/Nikhef, Utrecht, Netherlands\\
$^{59}$ Institute of Experimental Physics, Slovak Academy of Sciences, Ko\v{s}ice, Slovak Republic\\
$^{60}$ Institute of Physics, Homi Bhabha National Institute, Bhubaneswar, India\\
$^{61}$ Institute of Physics of the Czech Academy of Sciences, Prague, Czech Republic\\
$^{62}$ Institute of Space Science (ISS), Bucharest, Romania\\
$^{63}$ Institut f\"{u}r Kernphysik, Johann Wolfgang Goethe-Universit\"{a}t Frankfurt, Frankfurt, Germany\\
$^{64}$ Instituto de Ciencias Nucleares, Universidad Nacional Aut\'{o}noma de M\'{e}xico, Mexico City, Mexico\\
$^{65}$ Instituto de F\'{i}sica, Universidade Federal do Rio Grande do Sul (UFRGS), Porto Alegre, Brazil\\
$^{66}$ Instituto de F\'{\i}sica, Universidad Nacional Aut\'{o}noma de M\'{e}xico, Mexico City, Mexico\\
$^{67}$ iThemba LABS, National Research Foundation, Somerset West, South Africa\\
$^{68}$ Jeonbuk National University, Jeonju, Republic of Korea\\
$^{69}$ Johann-Wolfgang-Goethe Universit\"{a}t Frankfurt Institut f\"{u}r Informatik, Fachbereich Informatik und Mathematik, Frankfurt, Germany\\
$^{70}$ Korea Institute of Science and Technology Information, Daejeon, Republic of Korea\\
$^{71}$ KTO Karatay University, Konya, Turkey\\
$^{72}$ Laboratoire de Physique des 2 Infinis, Ir\`{e}ne Joliot-Curie, Orsay, France\\
$^{73}$ Laboratoire de Physique Subatomique et de Cosmologie, Universit\'{e} Grenoble-Alpes, CNRS-IN2P3, Grenoble, France\\
$^{74}$ Lawrence Berkeley National Laboratory, Berkeley, California, United States\\
$^{75}$ Lund University Department of Physics, Division of Particle Physics, Lund, Sweden\\
$^{76}$ Nagasaki Institute of Applied Science, Nagasaki, Japan\\
$^{77}$ Nara Women{'}s University (NWU), Nara, Japan\\
$^{78}$ National and Kapodistrian University of Athens, School of Science, Department of Physics , Athens, Greece\\
$^{79}$ National Centre for Nuclear Research, Warsaw, Poland\\
$^{80}$ National Institute of Science Education and Research, Homi Bhabha National Institute, Jatni, India\\
$^{81}$ National Nuclear Research Center, Baku, Azerbaijan\\
$^{82}$ National Research and Innovation Agency - BRIN, Jakarta, Indonesia\\
$^{83}$ Niels Bohr Institute, University of Copenhagen, Copenhagen, Denmark\\
$^{84}$ Nikhef, National institute for subatomic physics, Amsterdam, Netherlands\\
$^{85}$ Nuclear Physics Group, STFC Daresbury Laboratory, Daresbury, United Kingdom\\
$^{86}$ Nuclear Physics Institute of the Czech Academy of Sciences, Husinec-\v{R}e\v{z}, Czech Republic\\
$^{87}$ Oak Ridge National Laboratory, Oak Ridge, Tennessee, United States\\
$^{88}$ Ohio State University, Columbus, Ohio, United States\\
$^{89}$ Physics department, Faculty of science, University of Zagreb, Zagreb, Croatia\\
$^{90}$ Physics Department, Panjab University, Chandigarh, India\\
$^{91}$ Physics Department, University of Jammu, Jammu, India\\
$^{92}$ Physics Program and International Institute for Sustainability with Knotted Chiral Meta Matter (SKCM2), Hiroshima University, Hiroshima, Japan\\
$^{93}$ Physikalisches Institut, Eberhard-Karls-Universit\"{a}t T\"{u}bingen, T\"{u}bingen, Germany\\
$^{94}$ Physikalisches Institut, Ruprecht-Karls-Universit\"{a}t Heidelberg, Heidelberg, Germany\\
$^{95}$ Physik Department, Technische Universit\"{a}t M\"{u}nchen, Munich, Germany\\
$^{96}$ Politecnico di Bari and Sezione INFN, Bari, Italy\\
$^{97}$ Research Division and ExtreMe Matter Institute EMMI, GSI Helmholtzzentrum f\"ur Schwerionenforschung GmbH, Darmstadt, Germany\\
$^{98}$ Saga University, Saga, Japan\\
$^{99}$ Saha Institute of Nuclear Physics, Homi Bhabha National Institute, Kolkata, India\\
$^{100}$ School of Physics and Astronomy, University of Birmingham, Birmingham, United Kingdom\\
$^{101}$ Secci\'{o}n F\'{\i}sica, Departamento de Ciencias, Pontificia Universidad Cat\'{o}lica del Per\'{u}, Lima, Peru\\
$^{102}$ Stefan Meyer Institut f\"{u}r Subatomare Physik (SMI), Vienna, Austria\\
$^{103}$ SUBATECH, IMT Atlantique, Nantes Universit\'{e}, CNRS-IN2P3, Nantes, France\\
$^{104}$ Sungkyunkwan University, Suwon City, Republic of Korea\\
$^{105}$ Suranaree University of Technology, Nakhon Ratchasima, Thailand\\
$^{106}$ Technical University of Ko\v{s}ice, Ko\v{s}ice, Slovak Republic\\
$^{107}$ The Henryk Niewodniczanski Institute of Nuclear Physics, Polish Academy of Sciences, Cracow, Poland\\
$^{108}$ The University of Texas at Austin, Austin, Texas, United States\\
$^{109}$ Universidad Aut\'{o}noma de Sinaloa, Culiac\'{a}n, Mexico\\
$^{110}$ Universidade de S\~{a}o Paulo (USP), S\~{a}o Paulo, Brazil\\
$^{111}$ Universidade Estadual de Campinas (UNICAMP), Campinas, Brazil\\
$^{112}$ Universidade Federal do ABC, Santo Andre, Brazil\\
$^{113}$ University of Cape Town, Cape Town, South Africa\\
$^{114}$ University of Houston, Houston, Texas, United States\\
$^{115}$ University of Jyv\"{a}skyl\"{a}, Jyv\"{a}skyl\"{a}, Finland\\
$^{116}$ University of Kansas, Lawrence, Kansas, United States\\
$^{117}$ University of Liverpool, Liverpool, United Kingdom\\
$^{118}$ University of Science and Technology of China, Hefei, China\\
$^{119}$ University of South-Eastern Norway, Kongsberg, Norway\\
$^{120}$ University of Tennessee, Knoxville, Tennessee, United States\\
$^{121}$ University of the Witwatersrand, Johannesburg, South Africa\\
$^{122}$ University of Tokyo, Tokyo, Japan\\
$^{123}$ University of Tsukuba, Tsukuba, Japan\\
$^{124}$ University Politehnica of Bucharest, Bucharest, Romania\\
$^{125}$ Universit\'{e} Clermont Auvergne, CNRS/IN2P3, LPC, Clermont-Ferrand, France\\
$^{126}$ Universit\'{e} de Lyon, CNRS/IN2P3, Institut de Physique des 2 Infinis de Lyon, Lyon, France\\
$^{127}$ Universit\'{e} de Strasbourg, CNRS, IPHC UMR 7178, F-67000 Strasbourg, France, Strasbourg, France\\
$^{128}$ Universit\'{e} Paris-Saclay Centre d'Etudes de Saclay (CEA), IRFU, D\'{e}partment de Physique Nucl\'{e}aire (DPhN), Saclay, France\\
$^{129}$ Universit\`{a} degli Studi di Foggia, Foggia, Italy\\
$^{130}$ Universit\`{a} del Piemonte Orientale, Vercelli, Italy\\
$^{131}$ Universit\`{a} di Brescia, Brescia, Italy\\
$^{132}$ Variable Energy Cyclotron Centre, Homi Bhabha National Institute, Kolkata, India\\
$^{133}$ Warsaw University of Technology, Warsaw, Poland\\
$^{134}$ Wayne State University, Detroit, Michigan, United States\\
$^{135}$ Westf\"{a}lische Wilhelms-Universit\"{a}t M\"{u}nster, Institut f\"{u}r Kernphysik, M\"{u}nster, Germany\\
$^{136}$ Wigner Research Centre for Physics, Budapest, Hungary\\
$^{137}$ Yale University, New Haven, Connecticut, United States\\
$^{138}$ Yonsei University, Seoul, Republic of Korea\\
$^{139}$  Zentrum  f\"{u}r Technologie und Transfer (ZTT), Worms, Germany\\
$^{140}$ Affiliated with an institute covered by a cooperation agreement with CERN\\
$^{141}$ Affiliated with an international laboratory covered by a cooperation agreement with CERN.\\

\end{flushleft} 

\end{document}